\documentclass[usenatbib]{mn2e}
\usepackage[dvips]{graphicx}
\usepackage{amssymb,amsmath}
\usepackage[titletoc]{appendix}
\usepackage{times}
\usepackage{fleqn,subfigure,lscape,epsfig,graphics}
\usepackage[dvipsnames,usenames]{color}
\usepackage{colortbl}

\title[Radio-loud vs. radio-quiet: AGN activity and SFR]{$\textit{Herschel}$-ATLAS:The connection between star formation and AGN activity in radio-loud and radio-quiet active galaxies}
\author[G\"urkan et al.]
{
{\parbox{\textwidth}{G. G\"urkan$^{1}$\thanks{E-mail:g.gurkan-uygun@herts.ac.uk},
M.J. Hardcastle $^{1}$,
M.J. Jarvis $^{2,3}$,
D.J.B. Smith $^{1}$,
N. Bourne $^{4}$,\\
L. Dunne $^{4,5}$,
S. Maddox $^{4,5}$,
R.J. Ivison $^{4,6}$,
J. Fritz $^{7}$\vspace{0.4cm}\\}
}\\
$^{1}$ School of Physics, Astronomy and Mathematics, University of Hertfordshire, College Lane, Hatfield AL10 9AB, UK\\
$^{2}$ Astrophysics, Department of Physics, Keble Road, Oxford, OX1 3RH, UK\\
$^{3}$ Physics Department, University of the Western Cape, Private Bag X17, Bellville 7535, South Africa\\
$^{4}$ Institute for Astronomy, University of Edinburgh, Royal Observatory, Blackford Hill, Edinburgh, EH9 3HJ, UK\\
$^{5}$ Department of Physics and Astronomy, University of Canterbury, Private Bag 4800, Christchurch, 8140, New Zealand\\
$^{6}$ European Southern Observatory, Karl Schwarzschild Strasse 2, Garching bei Munchen, Germany\\
$^{7}$ Centro de Radioastronom\'\i a y Astrof\'\i sica, CRyA, UNAM, Campus Morelia, A.P. 3-72, C.P. 58089, Michoac\'an, Mexico\\
}

\begin{document}
% bibliography and bibfile journal definitions (taken from aa.cls)

% Astronomy and Astrophysics
\def\aj{AJ}					% Astronomical Journal
\def\araa{ARA\&A}				% Annual Reviews of Astronomy and Astrophysics
\def\nar{NewAR}                                 % New Astronomy Reviews
\def\apj{ApJ}					% Astrophysical Journal
\def\apjl{ApJL}					% Astrophysical Journal, Letters
\def\apjs{ApJS}					% Astrophysical Journal, Supplement Series
\def\apss{Astrophysics and Space Science}
\def\capsp{Comments on Astrophysics and Space Physics}
\def\aap{A\&A}					% Astronomy and Astrophysics
\def\aapr{A\&A~Rev.}				% Astronomy and Astrophysics Reviews
\def\aaps{A\&AS}				% Astronomy and Astrophysics, Supplement
\def\azh{AZh}					% Astronomicheskii Zhurnal
\def\baas{BAAS}					% Bulletin of the AAS
\def\jrasc{JRASC}				% Journal of the RAS of Canada
\def\memras{MmRAS}				% Memoirs of the RAS
\def\mnras{MNRAS}					% Monthly Notices of the Royal Astronomical Society
\def\pasp{PASP}					% Publications of the ASP
\def\pasj{PASJ}					% Publications of the ASJ
\def\qjras{QJRAS}				% Quarterly Journal of the RAS
\def\skytel{S\&T}				% Sky and Telescope
\def\solphys{Sol.~Phys.}			% Solar Physics
\def\sovast{Soviet~Ast.}			% Soviet Astronomy
\def\ssr{Space~Sci.~Rev.}			% Space Science Reviews
\def\zap{ZAp}					% Zeitschrift fuer Astrophysik
\def\na{New Astronomy}				% New Astronomy
\def\iaucirc{IAU~Circ.}				% IAU Cirulars
\def\aplett{Astrophys.~Lett.}			% Astrophysics Letters
\def\apspr{Astrophys.~Space~Phys.~Res.}		% Astrophysics Space Physics Research
\def\bain{Bull.~Astron.~Inst.~Netherlands}	% Bulletin Astronomical Institute of the Netherlands
\def\memsai{Mem.~Soc.~Astron.~Italiana}		% Mem. Societa Astronomica Italiana

% Optics
\def\ao{Appl.~Opt.}				% Applied Optics

% General physics
\def\pra{Phys.~Rev.~A}				% Physical Review A: General Physics
\def\prb{Phys.~Rev.~B}				% Physical Review B: Solid State
\def\prc{Phys.~Rev.~C}				% Physical Review C
\def\prd{Phys.~Rev.~D}				% Physical Review D
\def\pre{Phys.~Rev.~E}				% Physical Review E
\def\prl{Phys.~Rev.~Lett.}			% Physical Review Letters
\def\nat{Nature}				% Nature
\def\fcp{Fund.~Cosmic~Phys.}			% Fundamental Cosmic Physics
\def\gca{Geochim.~Cosmochim.~Acta}		% Geochimica Cosmochimica Acta
\def\grl{Geophys.~Res.~Lett.}			% Geophysics Research Letters
\def\jcp{J.~Chem.~Phys.}			% Journal of Chemical Physics
\def\jgr{J.~Geophys.~Res.}			% Journal of Geophysics Research
\def\jqsrt{J.~Quant.~Spec.~Radiat.~Transf.}	% Journal of Quantitiative Spectroscopy and Radiative Trasfer
\def\nphysa{Nucl.~Phys.~A}			% Nuclear Physics A
\def\physrep{Phys.~Rep.}			% Physics Reports
\def\physscr{Phys.~Scr}				% Physica Scripta
\def\planss{Planet.~Space~Sci.}			% Planetary Space Science
\def\procspie{Proc.~SPIE}			% Proceedings of the SPIE
\def\rpp{Rep.~Prog.~Phys.}			% Rep. Prog. Phys.
\let\astap=\aap
\let\apjlett=\apjl
\let\apjsupp=\apjs
\let\applopt=\ao
\let\prep=\physrep

% end of file

\date{Accepted ...... Received ...... ; in original form......   }

\pagerange{\pageref{firstpage}--\pageref{lastpage}} \pubyear{2011}
\maketitle
\label{firstpage}
\begin{abstract}

We examine the relationship between star formation and AGN activity by constructing matched samples of local ($0<z<0.6$) radio-loud and radio-quiet AGN in the $\textit{Herschel}$-ATLAS fields. Radio-loud AGN are classified as high-excitation and low-excitation radio galaxies (HERGs, LERGs) using their emission lines and $\textit{WISE}$ 22-$\mu$m luminosity. AGN accretion and jet powers in these active galaxies are traced by [OIII] emission-line and radio luminosity, respectively. Star formation rates (SFRs) and specific star formation rates (SSFRs) were derived using $\textit{Herschel}$ 250-$\mu$m luminosity and stellar mass measurements from the SDSS$-$MPA-JHU catalogue. In the past, star formation studies of AGN have mostly focused on high-redshift sources to observe the thermal dust emission that peaks in the far-infrared, which limited the samples to powerful objects. However, with $\textit{Herschel}$ we can expand this to low redshifts. Our stacking analyses show that SFRs and SSFRs of both radio-loud and radio-quiet AGN increase with increasing AGN power but that radio-loud AGN tend to have lower SFR. Additionally, radio-quiet AGN are found to have approximately an order of magnitude higher SSFRs than radio-loud AGN for a given level of AGN power. The difference between the star formation properties of radio-loud and -quiet AGN is also seen in samples matched in stellar mass.

\end{abstract}


\section{INTRODUCTION}
Since their discovery active galactic nuclei (AGN) have formed an important part of astrophysics research. Investigations of AGN are not only crucial in their own right, but also essential for galaxy formation and evolution studies. Accumulating observational data clearly show that the masses of black holes in massive galaxies are correlated with various properties of their hosts such as the galaxy luminosities \citep[e.g.][]{1995ref17,2003ref18,2009ref61}, the galaxy bulge masses \citep[e.g.][]{1998ref12,2002ref14} and the velocity dispersions \citep[e.g.][]{2000ref13,2000ref15,2001ref16}. Furthermore, the anti-hierarchical evolution of AGN \citep[e.g.][]{2000ref20,2005ref21,2007ref22,2011ref23}, i.e. the fact that  the space density of low-luminosity AGN peaks around $z<1$ and that of high-luminosity AGN peaks around $z\sim2$, is very similar to the cosmic downsizing of star forming galaxies \citep[e.g.][]{1996ref24,2008ref25,2009ref26} and spheroidal galaxies \citep[e.g.][]{2006ref27,2010ref28}. In addition to these, both the integrated cosmic star formation rate and the black hole accretion rate increase rapidly from $z\sim0$ out to $z\sim 2$ \cite[][]{2003ref30,2007ref158,2007ref29,2008ref31,2011ref23}. All these relationships indicate that the formation and growth of the black holes and their host galaxies are fundamentally intertwined. 

Although these relationships are observed, there is not yet a clear understanding of how black holes grow, the link between the growth of black holes and their host galaxy properties, and what leads to these connections. To explain the observed co-evolution of black holes and their hosts, interactions (AGN feedback) between the black hole at the centre of a galaxy and the gas and dust that it contains have been invoked in theoretical models \citep[e.g.][]{2004ref32,2005ref33,2006ref34,2006ref35,2009ref36,2011ref37}. Major mergers have been widely suggested as a triggering mechanism of AGN activity \citep[e.g.][]{2000ref42,2005ref43}. Secular processes (disk instabilities, minor mergers, recycled gas from dying stars, galaxy bars etc.) have also been discussed as a mechanism responsible for fuelling, in particular for low-luminosity AGN \citep[e.g.][]{2009ref44,2008ref45,2010ref46}. 

Models often represent the effects of feedback in two ways, denoted `quasar mode' and `radio mode' \citep[e.g.][]{2006ref34}. In quasar mode the energy release occurs as winds with sub-relativistic outflows and wide opening angle driven by the radiative output of AGN. The generated radiation interacts with the gas and dust in the host galaxies and the resulting winds (either energy driven winds or momentum driven winds, see \cite{2009ref47} for further information) can expel the gas from the galaxy. This can stop the accretion of matter onto a black hole and further quench the formation of stars \citep[e.g.][]{2012ref53}. However, studies of X-ray luminous AGN do not show any evidence for this \citep[e.g.][but see \citealt{2012ref173}, \citealt{2014ref162} and \citealt{2015ref161}]{2012ref54}.

In radio-mode feedback, the accretion of the matter does not lead to powerful radiative output; instead we see the production of highly energetic jets. The jets may play an important role in the fate of the host galaxy by heating up the cold gas and suppressing star formation \citep[e.g.][]{2005ref57,2013ref9}. There is even an indication for this in radio-quiet quasars \citep[e.g.][]{2015ref178}. In the most dramatic scenario, the jets can expel the molecular gas from the host galaxy and stop star formation. Radio mode feedback has been widely used in simulations as a mechanism to prevent the overproduction of stars (by shutting down the star formation) in massive galaxies in order to produce observed ``red and dead'' early-type galaxies \citep[e.g.][]{2006ref60,2013ref58,2013ref59}. However, a positive radio mode feedback has also been suggested \citep{2010ref62,2012ref63,2014ref143,2012ref64}. In this scenario, the radio jets drive shocks in the interstellar medium which enhance the star formation, something that has been observed for decades now \citep[e.g.][]{1990ref154,1998ref155,2001ref156,2005ref157}.

Different types of AGN might be able to provide feedback in multiple ways. It is known that low-excitation radio galaxies (LERGs) do not have radiatively efficient accretion, which removes the possibility of quasar-mode feedback. They are believed to be fuelled by advection-dominated accretion flows \citep[e.g.][ADAFs]{1995ref74} which can create an environment where the energy release occurs kinetically by the radio jets. These jets are able to provide `radio-mode' feedback. Radio-quiet active galaxies and radio-quiet quasars have typical AGN properties where we see the radiative output but no strong radio jets, and so we can only expect to see `quasar mode' feedback. On the other hand, radio-loud quasars and HERGs have radiative output produced by highly efficient accretion as well as kinetic energy release seen as strong radio jets. Therefore, both feedback mechanisms might be expected to be observed in these powerful objects.

The anticipated relationship between AGN luminosity (or black hole accretion rate) and SFR has been investigated many times previously with mixed results (For recent reviews on black holes and galaxy evolution see \citealt{2012ref76}, \citealt{2013ref114} and \citealt{2014ref138}). For instance, some studies have found a slight correlation between these quantities \citep[e.g.][]{2008ref90,2009ref91,2011ref92,2009ref94,2013ref108}, some others found a strong correlation \citep[e.g.][]{2005ref93,2005ref107,2009ref95,2009ref96,2013ref97} whereas others \cite[e.g.][]{2012ref54,2012ref98,2012ref99,2012ref52,2012ref109} found a flat relationship (or no evidence for a correlation). On the other hand, the results of \cite{2012ref53} indicate a $\textit{suppression}$ of star formation due to AGN feedback (see \citealt{2014ref100} for a recent review on the IR perspective of AGN). This wide variety of results is puzzling. It is important to note that although there are some overlaps between samples, fields and indicators used for SFR and AGN luminosity, in general both the sample selection and the star-formation indicator used varies from study to study. These may cause biases in the conclusions derived. Thus, performing complete and coherent surveys of AGN, minimising the systematic uncertainties, and proposing revised models and testing them are crucial for future AGN research. Additionally, it has been pointed out that the different variability time scale of AGN activity and star formation can lead to these different results \citep[e.g.][]{2014ref125,2010ref140,2009ref129}. For this reason, instead of concentrating on individual sources and instantaneous AGN activity, averaging the AGN luminosity over the populations should provide a clearer view of the relation between SFR and AGN activity \citep[e.g.][]{2013ref97}.

Another interesting aspect is to search for differences in the relationship between the AGN activity and SFR for radio-loud and radio-quiet AGN. This has been investigated previously: \cite{2012ref103} found that at low redshifts radio-quiet AGN hosts have stars forming at higher rates than radio-loud counterparts while \cite{2013ref59} observed two times more actively star-forming galaxies among radio-quiet AGN than radio-loud AGN for galaxies with stellar masses M$_{*}>10^{11.4}$ M$_{\odot}$. These studies have provided important information on investigations of AGN$-$SFR relation for different types of AGN. However, it is still important to investigate similarities or differences between the relationship of AGN outputs (kinetic or radiative) and SFR using radio-quiet and radio-loud AGN samples matched in AGN power and stellar mass, and the possible reasons for these.

In this paper we investigate the role of AGN activity in regulating the host galaxy evolution as a function of the different types of AGN. In the pre-$\textit{Herschel}$ era, star-formation studies of radio-loud AGN concentrated on high-redshift sources in order to observe the thermal dust emission peaking in the far-infrared/sub-mm, which limited the samples to powerful objects \citep[e.g.][]{2001ref65,2004ref66}. However, it is now possible with $\textit{Herschel}$ to expand this to low redshifts \citep[e.g.][]{2010ref7,2010ref67,2013ref9,2013ref68}. Therefore, we are able to create matched samples of radio-loud and radio-quiet AGN with redshifts $0<z<0.6$ in terms of the relationship between their star formation properties and their AGN activity.

The layout of this paper is as follows. A description of the sample and, the classification of the AGN are given in Section 2. Our key results are given in Section 3, where the comparison of the stacking analysis between radio-loud and radio-quiet AGN as well as between HERGs and LERGs are presented. We also form a sample of sources matched in their stellar mass in order to investigate the star formation properties of these sources by excluding the effect of mass. Additionally the relation between SFR and black hole accretion rate ($\dot{M}_{\textrm{BH}}$) for radio-loud and radio-quiet AGN samples is examined. In Section 4 we interpret our results. Section 5 presents a summary of our results and conclusions.

The cosmological parameters used throughout the paper are as follows: $\Omega_{m}$=0.3, $\Omega_{\Lambda}$=0.7 and $H_{0}$=70 km s$^{-1}$ Mpc$^{-1}$. 
%%%%%%%%%%%%%%%%%%%%%%%%%%%%%%%%%%%%%%%%%%%%%%%%%%%%%%%%%%%%%%%%%%%%%%%%%%%%%%%%%%%%%%%

\section{DATA}

 \subsection{Sample and classification}
 To construct our sample we selected galaxies from the seventh data release of the Sloan Digital Sky Survey \citep[SDSS; ][]{2009ref113} catalogue with the value-added spectroscopic measurements produced by the group from the Max Planck Institute for Astrophysics, and the John Hopkins University (MPA-JHU)\footnote{http://www.mpa-garching.mpg.de/SDSS/}. It has 31001 sources spanning a redshift range $0<z<0.7$. The sample does not include quasars and Type-I Seyferts because the AGN outshine the host galaxies for these objects which makes it difficult to study their host galaxy properties. The Herschel Astrophysical Terahertz Large Area Survey \citep[H-ATLAS; ][]{2010ref1} North Galactic Pole (NGP) and the three equatorial Galaxy and Mass Assembly (GAMA) fields \citep{2011ref144} data were used to obtain far-IR fluxes of the sample galaxies. Stellar mass estimates are available for most of the sources in the catalogue \citep{2003ref116}.

\cite{2012ref3} (BH12 hereafter) constructed a radio-loud AGN sample by combining the seventh data release of the SDSS sample with the National Radio Astronomy Observatory (NRAO) Very Large Array (VLA) Sky Survey (NVSS; \citep{1998ref117}) and the Faint Images of the Radio Sky at Twenty centimetres (FIRST) survey \citep{1995ref118} following the methods described by \cite{2005ref57} and \cite{2009ref169}. We will briefly summarize their method, further details are given by \cite{2005ref57} and \cite{2009ref169}. Firstly, each SDSS source was checked to see whether it has a NVSS counterpart: in the case of multiple-NVSS-component matches the integrated flux densities were summed to obtain the flux density of a radio source. If there was a single NVSS match, then FIRST counterparts of the source were checked. If a single FIRST component was matched, accepting or rejecting the match was decided based on FIRST properties of these sources. If there were multiple FIRST components  the source was accepted or rejected based on its NVSS properties. Since this classification is sensible and re-producible we use the classification of BH12 for our work here.

\begin{figure}
\begin{center}
\scalebox{0.9}{
\begin{tabular}{c}
\centerline{\hspace{-0.9em}\includegraphics[width=10.0cm,height=10.0cm,angle=0,keepaspectratio]{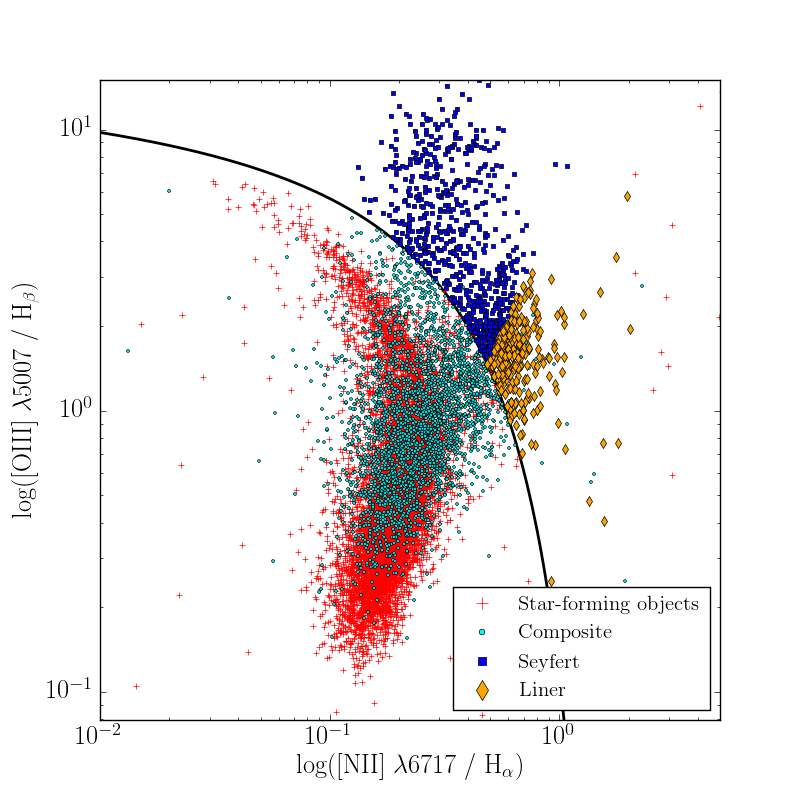}}\\
\end{tabular}}
\caption[.]{In order to present the clear separation between Seyferts and Liners we show here the [SII]/[H$_{\alpha}$]$-$[OIII]/[H$_{\beta}$] diagnostic diagram for galaxies in our sample, excluding sources from the radio-loud sample. A solid line represents the main AGN/SF division line summarized by \cite{2006ref2}.\label{classification} }
\end{center}
\end{figure}

We firstly cross-matched the initial sample of galaxies with the BH12 catalogue, and all objects that they classified as AGN form our `radio-loud' subsample, which has 613 objects. The remaining `radio-quiet' AGN sample are classified using the modified emission-line diagnostics given by \cite{2006ref2}. This classification is shown in Figure \ref{classification}. Composite objects were separated from star-forming objects using a classification line given by \cite{2003ref88}. This classification line utilizes the [NII] $\lambda$6584 / [H$_{\alpha}$] ratio so it is not shown in Fig. \ref{classification}.

\begin{figure}
\begin{center}
\scalebox{0.85}{
\begin{tabular}{c}
\centerline{\hspace{-0.9em}\includegraphics[width=11cm,height=11cm,angle=0,keepaspectratio]{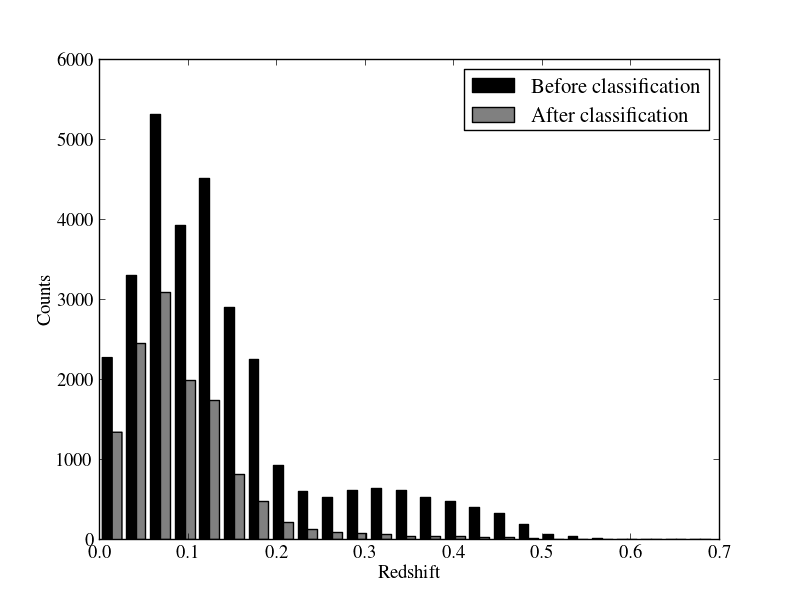}}\\
\centerline{\hspace{-0.9em}\includegraphics[width=11cm,height=11cm,angle=0,keepaspectratio]{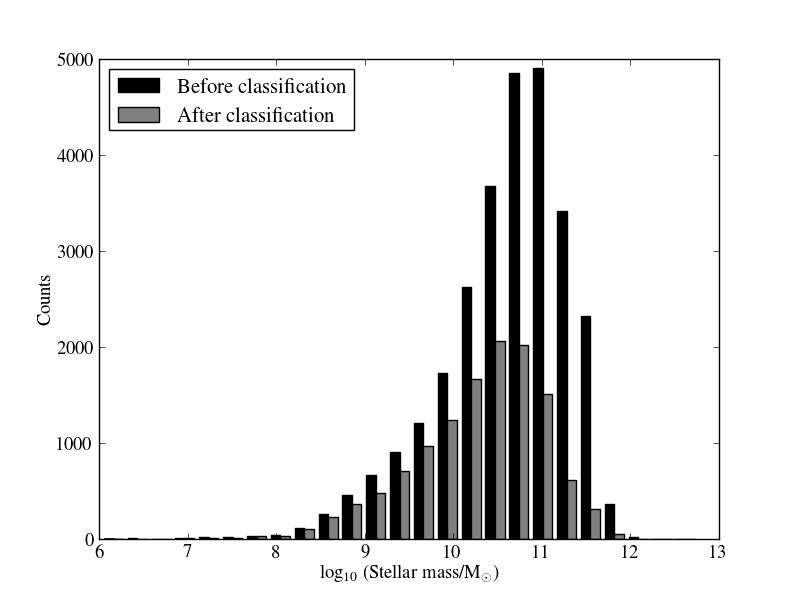}}\\
\end{tabular}}
\caption[.]{Redshift (top panel) and stellar mass (lower panel) distribution of the whole sample before implementing the emission-line classification (black) and the distribution of the same properties of the sample after the classification process (grey). \label{z_dist} }
\end{center}
\end{figure}

The classification done by using optical emission line ratios and this process gave us 8035 star-forming objects, 1190 objects classified as Seyferts, 2490 composite objects and 319 LINERs (Table \ref{sample-table}). Emission lines were not detected for 17741 objects in the sample so they could not be classified in this way. This classification biases us towards massive, low redshift galaxies. The sample has a mean stellar mass, log$_{10}$$(M/M_{\odot}$) = 9.55 and a mean redshift = 0.14. After the classification using optical emission lines the mean stellar mass is log$_{10}$($M/M_{\odot}$) = 10.08  and the mean redshift equals 0.09 (Figure \ref{z_dist} - bottom plot). The final sample has a redshift range $0<z<0.56$. We do not consider objects that are not classified using the BPT diagram (Fig. \ref{classification}) further in the paper, and we also discarded: star-forming objects, LINERs and the objects in the BH12 sample classified as star-forming sources from our AGN sample. The end result is that a combination of Seyferts and composite objects classified using optical emission lines formed our radio-quiet AGN sample, which has 3680 objects.

\begin{table}
\begin{tabular}{cccccccccccc}
\hline
&&&&&Population type&Counts&&&&&\\
\hline
&&&&&Radio-loud AGN&613&&&&&\\
&&&&&Seyferts&1190&&&&&\\
&&&&&Composites&2490&&&&&\\
&&&&&Star-forming objects&8035&&&&&\\
&&&&&LINERs&319&&&&&\\
\hline
\end{tabular}
\caption[]{The number of sources in each population after the optical emission-line classification.}{\label{sample-table}}
\end{table}

  The treatment of composite objects is discussed in detail in Section 3.1. For our analysis we use stellar mass estimations of the sources to be able to calculate their SSFRs. There are only 206 objects in the sample do not have these measurements. These sources were excluded from the parts of the analysis that involve SSFRs. These sources are systematically at higher redshifts so the sample used in the parts of analysis that involve SSFRs will be biased.

To be able to evaluate the properties of radio-loud AGN as a function of emission-line class, the radio galaxies in the sample were classified as high- and low-excitation radio galaxies (HERGs and LERGs). We initially used classification information provided by BH12. There are 191 sources which are not classified either as HERG or LERG by BH12. For these sources we used the \textit{WISE}-based (details with regard to $\textit{WISE}$ data are given in Section 2.3) classification proposed by \cite{2014ref4}. Our final radio-loud AGN sample has 404 LERGs and 209 as HERGs.

There are various diagnostics using radio and optical measurements to classify sources as radio-loud or radio-quiet AGN. We did not use any of these diagnostics to classify our objects as radio-loud or radio-quiet AGN. This is because our objects are Seyferts 2s and radio galaxies and the estimation of optical luminosities for such objects, even when they are radiatively efficient, is difficult. However, we checked whether our radio-loud and radio-quiet AGN classifications coincide with traditional classifications. In order to do this assessment, we used radio and emission-line measurements available to us. We calculated the ratio of radio luminosity (at 1.4 GHz) to an estimate of optical luminosity (at 5100 \AA). As our AGN are mostly Type II Seyferts, we do not know their intrinsic optical luminosities and to derive these we used our bolometric luminosities ($L_{bol}$, see Section 2.4) and the relation between $L_{bol}$ and $L_{5100}$ given by \cite{2012ref153}. We have 1.4 GHz fluxes of the radio-loud sources from BH12. We cross-matched the catalogue of our radio-quiet AGN sample to the FIRST catalogue to obtain their 1.4 GHz fluxes. This provided 131/1190 Seyfert detections and 211/2490 detections of composites. For the sources that are not detected in FIRST we used the measured minimum flux in the catalogue as an upper limit. We assumed a spectral index 0.8 (S$_{\nu}$ $\propto$ $\nu^{-\alpha}$) which is a typical expected value for radio sources \citep[e.g.][]{1964ref176,2010ref7} and derived the 1.4 GHz radio luminosities using FIRST fluxes of the sources. 

\begin{figure}
\begin{center}
        \resizebox{1.06\hsize}{!}{\hspace{-2.9em}\includegraphics{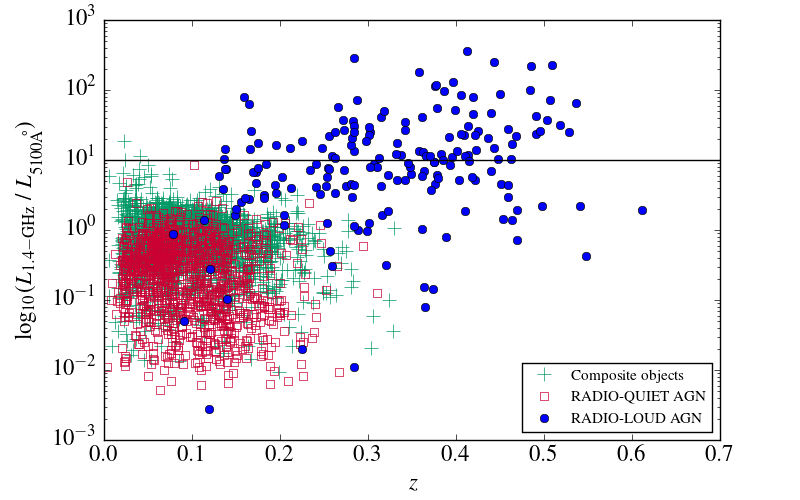}}\\
        %\resizebox{1.05\hsize}{!}{\includegraphics{MASS-PWR-L-Q-NO-COMPOSITE.png}}\\
        %\resizebox{1.05\hsize}{!}{\includegraphics{SSFR-PWR-L-Q-NO-COMPOSITE.png}}\\
        \caption{The distribution of the radio-loudness parameter for the sample of radio-quiet AGN and HERGs used in this work. LERGs are not shown as they are not radiatively efficient AGN. We used an estimate of optical luminosity at 5100 \AA\,and 1.4-GHz radio luminosity to calculate the parameter. \label{rl-rq-class}}
\end{center}\vspace{-1em}
\end{figure}

We exclude LERGs from the analysis because they do not have significant AGN-related optical continuum emission. The results of this analysis are plotted in Fig. \ref{rl-rq-class}. We see that there is a clear separation between the radio-loud and radio-quiet samples using this parameter, in the sense that the vast majority of objects that we have classified as radio-loud following BH12 have $r = \log_{10}(L_{\rm radio}/L_{\rm optical})>0.5$, while almost none of the radio-quiet objects exceed this value. It is important to note that traditionally a threshold $r>1$ is used to select radio-loud objects (solid line on Fig. \ref{rl-rq-class}) adopting this threshold would classify some of our radio sources as radio-quiet. However, (i) this is an essentially arbitrary dividing line, and (ii), given the systematic uncertainties in computing $r$ for this sample introduced by the use of bolometric corrections, we cannot really be certain that the discrepancy between our observed threshold in $r$ and the traditional one is significant. Reclassifying the very few radio-loud sources with $r\ll 0.5$ as radio-quiet would not affect our analysis in any way. Given the broad consistency between the two radio-loudness estimates we choose to retain our original classification based on the BH12 analysis, which is easily reproducible by future workers. We also calculated two different radio loudness parameters which are commonly used: the ratio of 5-GHz flux density to the flux density at 2500 \AA\,\citep[e.g.][]{1992ref174} and the ratio of 5-GHz flux density to the flux density at 4400 \AA\,\citep[e.g.][]{1989ref175}. 5-GHz radio flux densities were extrapolated assuming a spectral index 0.8. In the same way, we used the relations given by \cite{2012ref153} between $L_{bol}$ and $L_{5100}$, and $L_{bol}$ and $L_{3000}$ to estimate fluxes at 5100 \AA\,and at 3000 \AA, respectively. Considering our sources are at low redshift these values will be approximately equal to the fluxes at 4400 \AA\,and 2500 \AA. By comparing these parameter estimations with the example shown in Fig. \ref{rl-rq-class} we found that there is an agreement between different radio-loudness ratios in terms of radio-quiet AGN having ratios lower than the given parameter value. For all the reasons mentioned above we did not use any of these parameters to classify our sources. All of our estimations are provided in the online table. We can also see from this figure that the redshift range is higher for radio-loud objects than radio-quiet sources. This is because there is not enough volume at low redshift coupled with the steep evolution of radio sources. Additionally, radio sources' host galaxies, which tend to be massive ellipticals, can be easily observed by SDSS to higher redshifts.

 \subsection{Far-infrared data, star formation and specific star formation rates}

$\textit{Herschel}$-ATLAS provides imaging data for $\sim$ 600 square degrees using the Photo-detector Array Camera and Spectrometer \citep[PACS; at 100-, 160-$\mu$m][]{2010ref10,2010ref166} and the Spectral and Photometric Imaging Receiver \citep[SPIRE; at 250-, 350- and 500-$\mu$m,][Valiante et. al. in prep.]{2010ref11,2011ref145}. To derive the maximum-likelihood estimate of the flux densities at the positions of objects in the SPIRE bands the flux density from the point spread function (PSF)-convolved H-ATLAS images were measured for each source together with the error on the fluxes. Further details about flux measurements can be found in \cite{2010ref7,2013ref9}.

Far-IR luminosity is widely used to measure star formation activity. As discussed by \cite{2013ref9} neither the integrated far-IR luminosity nor the 250-$\mu$m luminosity is a completely reliable estimator of SFR because of the contribution by cold dust heated by old stellar populations \citep{2010ref146,2012ref147}. However, we are secure using them as long as the far-IR emission is dominated by warm dust heated by star formation which, gives rise to dust temperatures $\sim$25 K or more. We utilised 250-$\mu$m luminosity, as the star-formation indicator that is least affected by dust heated by the AGN \citep[e.g.][]{2010ref126,2010ref7,2012ref127}. For the sources with PACS and SPIRE detections we used a modified black-body spectrum for the far-IR spectral energy distribution (SED; $f_{\nu}$ $\propto$ $\frac{\nu^{3+\beta}}{e^{\frac{h\nu}{kT}}-1}$); we fixed the emissivity index $\beta$ as 1.5 and allowed temperatures to vary to obtain the best fitting temperatures, integrated luminosities ($L_{\mathrm{IR}}$) and rest-frame luminosities at 250-$\mu$m ($L_{250}$) provided by the minimum $\chi^{2}$ values. To calculate the $K$-corrections the same emissivity index ($\beta$=1.5) and the mean of the best-fitting temperatures for each class were used (See Table \ref{temp-table} for the temperatures). These corrections were included in the derivation of 250-$\mu$m luminosities.

Table \ref{temp-table} shows the best-fitting temperatures for each type of AGN in the sample, together with uncertainties derived by the bootstrap method. The results show that the mean temperatures of the populations in the sample are around 25 K (Identical temperature estimates for star-forming galaxies were also reported by \citealt{2013ref87}). This indicates that the far-IR measurements of our sources are not strongly affected by cold dust heated by the old stellar populations. It has been suggested that SFRs derived from far-IR luminosity overestimate SFRs of the sources with SFR below $\sim$1 M$_{\odot}$ yr$^{-1}$ due to the contribution by old stars \citep[e.g.][]{2013ref165}. However, most of our sources in the sample have SFRs above 1  M$_{\odot}$ yr$^{-1}$ (Fig. \ref{mass-sfr}) and any possible contribution by old stellar population should not be a problem for the SFR estimates.

Another source of contamination that we may observe at far-IR wavelengths is synchrotron emission from the jets of a radio-loud AGN. In order to check any possible synchrotron contamination in the $\textit{Herschel}$ 250-$\mu$m band we cross matched our radio-loud AGN objects to the Giant Metre-wave Radio Telescope (GMRT) catalogue by following the method described by \cite{2013ref141}. This provided 325-MHz GMRT fluxes for 165 out of 613 objects. The reason for this low success is that the GMRT survey only covers 36 per cent of the GAMA and NGP fields, and has incomplete sky coverage and variable sensitivity. Assuming that this sub-sample is representative of the whole sample, spectral indices of the matched objects, derived using the NVSS and GMRT fluxes, were used to gain extrapolated fluxes at 250 $\mu$m. Comparison of these with 250-$\mu$m far-IR fluxes showed that 100/165 objects have steep spectra and accordingly have much lower extrapolated flux than the measured far-IR flux at 250-$\mu$m. 65 out of 165 sources have extrapolated fluxes higher than their far-IR fluxes at 250 $\mu$m, but of these, almost all (59) are not detected at 250-$\mu$m $\textit{Herschel}$ band. This demonstrates that the synchrotron contamination in the far-IR is not a serious issue for our sample.

\begin{table}
\begin{tabular}{ccccccccc}
\hline
&&&Population type&Temperature (K)&$\sigma$ (K)&&&\\
\hline
&&&Radio-loud AGN&19.6&1.06&&&\\
&&&Radio-quiet AGN&25.6&0.37&&&\\
&&&Composites&27.2&0.23&&&\\
&&&Star-forming objects&25.8&0.12&&&\\
\hline
\end{tabular}
\caption[]{The mean temperature values for different populations with errors on them derived by the bootstrap technique.}{\label{temp-table}}
\end{table}

\begin{figure}
\begin{center}
\scalebox{0.89}{
\begin{tabular}{c}
\centerline{\hspace{-0.9em}\includegraphics[width=11cm,height=11cm,angle=0,keepaspectratio]{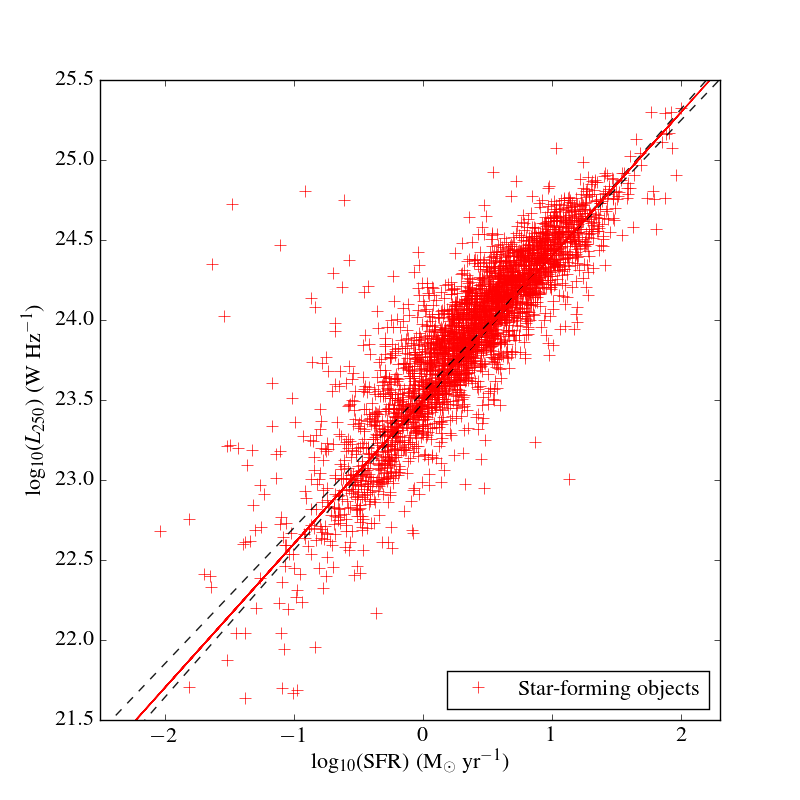}}\\
\end{tabular}}
\caption[.]{The distribution of the luminosity at 250-$\mu$m of the star-forming objects as a function of their SFRs. The best fit between the quantities is presented by a red line and $\pm \sigma$ errors of this fit are shown with grey dash lines. \label{correlation} }
\end{center}
\end{figure}

SFR measurements for galaxies in the SDSS 7th data release catalogue, calculated using the H$_{\alpha}$ emission line, corrected for dust attenuation and fiber aperture effects, can be found in the MPA-JHU data base \citep{2004ref8}. To determine the SFR/L$_{250}$ relation, we chose star-forming objects, classified using emission lines (Section 2.1), and which have both $L_{250}$ and SFR estimates. We then used the median likelihood SFR estimates derived using the H$_{\alpha}$ emission line given in the catalogue to calibrate the relationship between SFR and $L_{250}$\footnote{It would be desirable to use additional data to estimate the true SFRs of galaxies taking into account the unobscured fraction that is not observed at far-IR bands. However, by calibrating our SFR relation using SFR determined using the H$_{\alpha}$ emission line, we take into account the unobscured fraction in our estimates.}. The errors in SFRs were estimated using the 16th and 84th percentiles.  We then used a Markov-Chain Monte Carlo regression to obtain the relation between SFR and $L_{250}$, considering the errors on both SFRs and $L_{250}$, and incorporating an intrinsic dispersion in the manner described by \cite{2010ref7}. The derived Bayesian estimate of the slope and intercept of the correlation are\\
\\
%log$_{10}$($L_{IR}$/L$_{\odot}$) = 10.1$(\pm0.01)$ + 0.96($\pm$0.02)log$_{10}$(SFR/$M_{\odot}$ yr$^{-1}$)\\
log$_{10}$($L_{250}$/W Hz$^{-1}$)\:=\:23.5$(\pm0.01)$\:+\:0.90$(\pm0.02)$log$_{10}$(SFR/$M_{\odot}$ yr$^{-1}$).\\
\\ 
 The slope shown as the red line is very close to unity and the derived relation is in agreement with the results of \cite{2013ref9}. The relationship between SFR and $L_{250}$ is presented in Figure \ref{correlation}. The relation between SFR and the integrated 8-1000 $\mu$m far-IR luminosity was also verified and it is found to be in agreement with the results reported by \cite{1998ref86} and \cite{2011ref164}.

For a further evaluation of $L_{250}$ as a star-formation indicator we compared the median-likelihood SFR estimates derived from Multi-wavelength Analysis of Galaxy Physical Properties (Magphys) SED fitting \citep{2008ref167,2012ref147}, where cold dust and unobscured far-IR emission are taken into account, and the SFR estimates from $L_{250}$. This comparison showed that SFRs derived from these different methods are consistent with each other once the uncertainties are taken into account. In particular, we found that the slope of a regression line between the two SFR estimates, and the mean ratio of SFR$_{L250}$ to SFR$_{Magphys}$, are both consistent with unity.

 \subsection{Near- and mid-infrared data}
Mid-IR luminosity can be used as an AGN power indicator because emission due to accretion is obscured by dust and gas structure (if it is present) and re-radiated in the mid-IR. The $\textit{WISE}$ mission has observed the whole sky in four mid-IR bands ($W1$ [3.4-$\mu$m], $W2$ [4.6-$\mu$m], $W3$ [12-$\mu$m], $W4$ [22-$\mu$m]) with an angular resolution of 6.1, 6.4, 6.5 and 12 arcsec, respectively. The $\textit{WISE}$ all-sky catalogue was searched for all objects in our samples. This was done by searching the catalogue within 3 arcsec from the coordinates of the sample sources. 30416 sources of the sample before classification had $\textit{WISE}$ detections. The $\textit{WISE}$ measurements are given in Vega magnitudes so the magnitudes of the sources were converted into Jy using the standard $\textit{WISE}$ zero-points\footnote{The relation used for the conversion can be found at http://wise2.ipac.caltech.edu/docs/release/allsky/expsup/}.

 In order to compute the luminosity at 22-$\mu$m, 12-$\mu$m$-$22-$\mu$m, spectral indices ($F_{\nu}$ $\alpha$ $\nu^{-\alpha}$ assuming a power law slope for the spectral energy distributions-SED) were calculated based on the $\textit{WISE}$ photometry for detected sources. K-corrections were not derived using SEDs because different components (old stellar population, torus and star formation) emit in the mid-IR.  In order to be consistent we derived the index to be used for $K$-corrections by taking the mean value of the indices of all sources. The mean index is 2.26 ($\sigma$=1.08). 
 
\subsection{AGN power indicators}
The 22-$\mu$m and [OIII] $\lambda 5007$ fluxes and luminosities are available to us, and can both be used as AGN power indicators. In order to see from where the mid-IR emission originates we carried out Markov-Chain Monte Carlo regression analysis to obtain the relation between the $L_{250}$ (W Hz$^{-1}$), as obtained from the temperature fitting, and the $\nu L_{\nu}$ luminosity at 22-$\mu$m ($L_{\mathrm{22}}$ erg s$^{-1}$). For this analysis we used the star-forming objects that were detected in both $\textit{Herschel}$ 250-$\mu$m and $\textit{WISE}$ 22-$\mu$m bands. This provided the following relation:\\
\\
log$_{10}$($L_{\mathrm{22}}$/erg s$^{-1}$) = $-7.93$ + 1.18 ($\pm$0.04) log$_{10}$($L_{250}$/erg s$^{-1}$).\\
\\

\begin{figure*}
\begin{center}
        \resizebox{0.99\hsize}{!}{\includegraphics{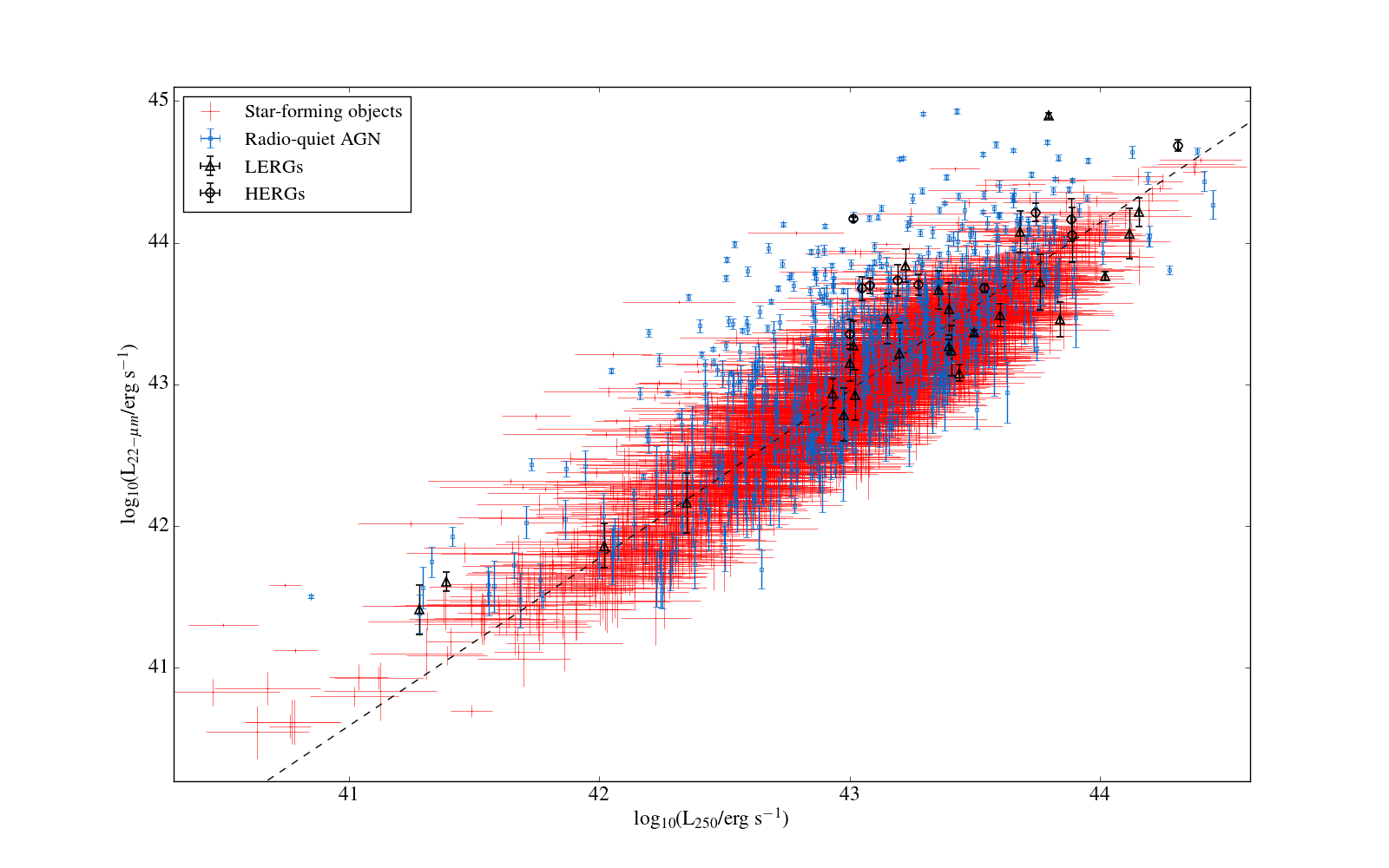}}\\
        \resizebox{0.497\hsize}{!}{\includegraphics{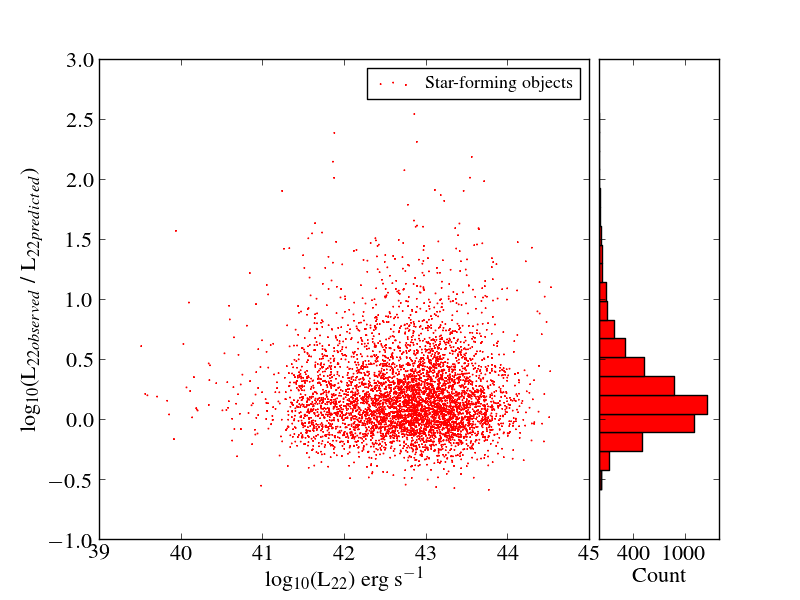}}
        \resizebox{0.497\hsize}{!}{\includegraphics{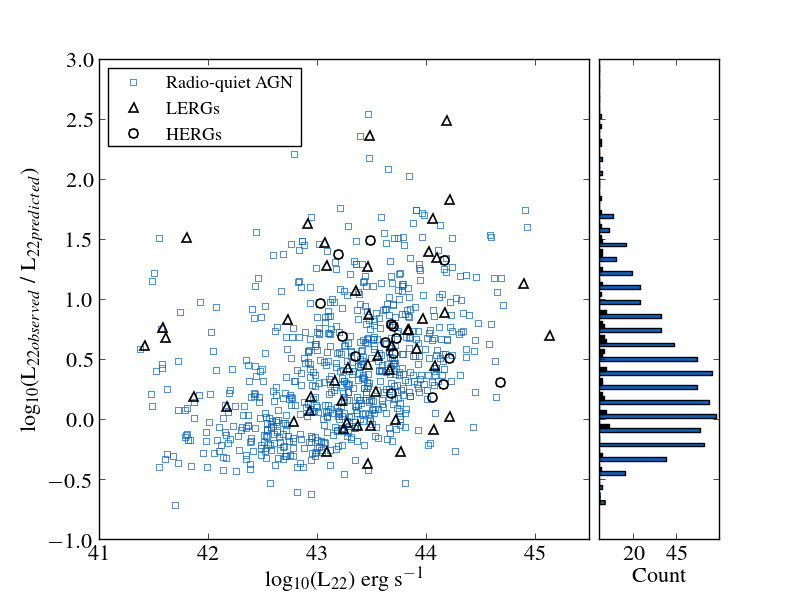}}
        \caption{The top panel shows the relation between L$_{22}$ and L$_{250}$ for star-forming objects and our sample sources that are detected in $\textit{WISE}$-22-$\mu$m band. In the bottom the distribution of the residual luminosities of the star-forming objects (left) and AGN in our sample detected at 22 $\mu$m (right) as a function of $L_{\mathrm{22}}$  together with a histogram of the counts are seen. The dashed line on the top panel shows the best fit for star forming objects. Star-forming objects are plotted as red crosses, radio-quiet AGN as blue squares, HERGs as open black circles and LERGs as open black triangles. }
\label{residual1}
\end{center}\vspace{-1em}
\end{figure*}

We then used this relation to derive the expected 22-$\mu$m luminosities due to star formation alone of all objects in the sample and the star-forming objects. In order to understand the effect of star formation on the 22-$\mu$m $\textit{WISE}$ band we subtracted these estimations from the actual 22-$\mu$m luminosity measurements. These residuals were then plotted as a function of mid-IR luminosity. In Figure \ref{residual1} we show the results of this analysis. In the top plot the relationship between log$_{10}$($L_{22}$) and log$_{10}$($L_{250}$) of the sources in our sample is seen, and we can see that there is a close relationship between log$_{10}$($L_{\mathrm{22}}$) and log$_{10}$($L_{250}$). It is also apparent that radio-quiet and radio-loud sources have similar distributions to star-forming objects. The bottom plots present the residual 22-$\mu$m luminosities of star-forming objects (left) and sources of the radio-loud and radio-quiet AGN sample (right) as function of $L_{\mathrm{22}}$ (erg s$^{-1}$). The comparison of these clearly shows that the mid-IR emission is dominated by star formation even for the AGN. As expected, star-forming objects cluster around the residual luminosity of 0. 

We implemented the same process using $L_{\mathrm{[OIII]}}$ for a comparison, obtaining the relation for pure star-forming objects as follows:\\
\\ 
log$_{10}$( $L_{\mathrm{[OIII]}}$/erg s$^{-1}$) = $-10.93$ + 1.16 ($\pm$0.05) log$_{10}$($L_{250}$/erg s$^{-1}$).\\
\\
 
\begin{figure*}
\begin{center}
        \resizebox{0.99\hsize}{!}{\includegraphics{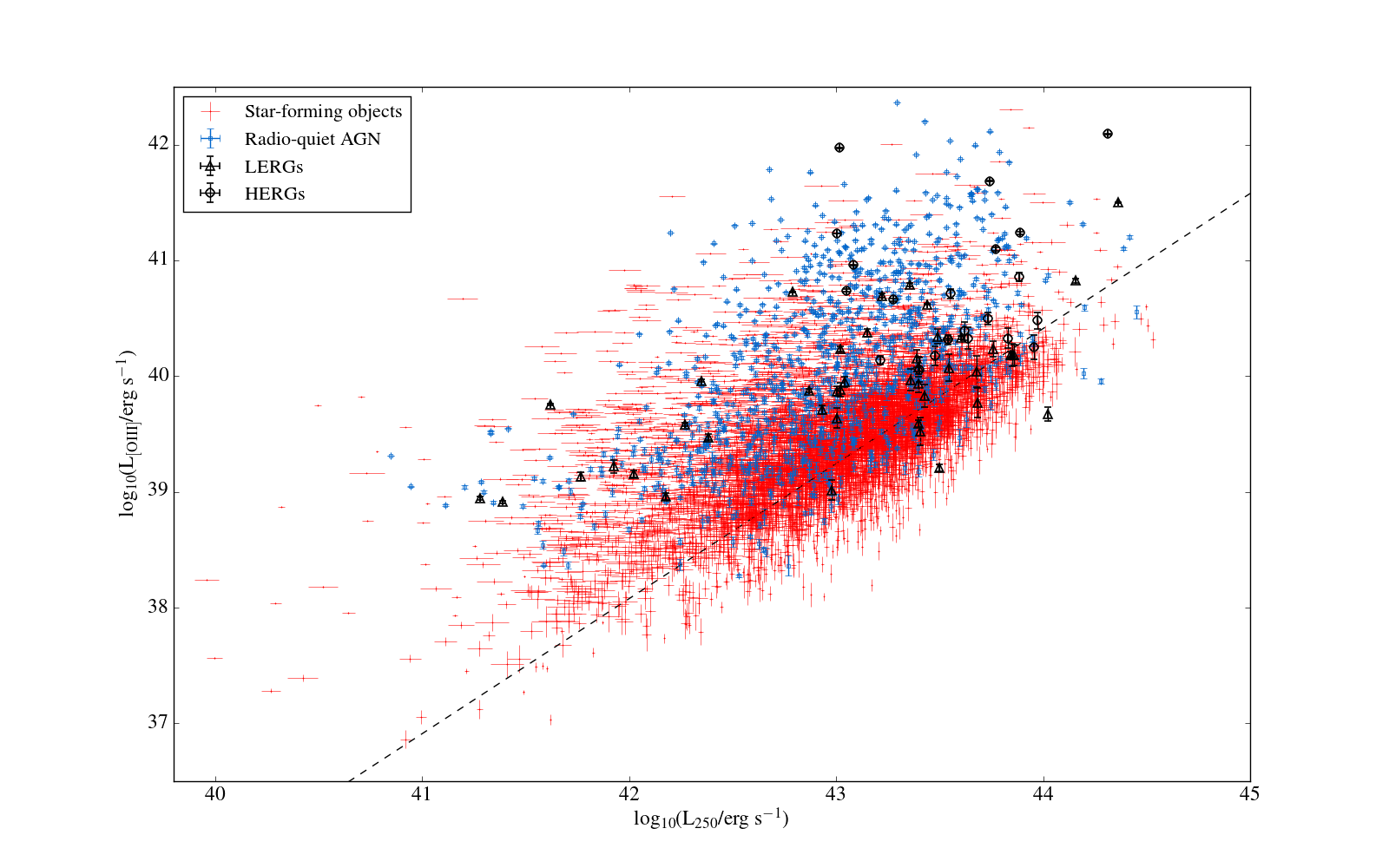}}\\
        \resizebox{0.497\hsize}{!}{\includegraphics{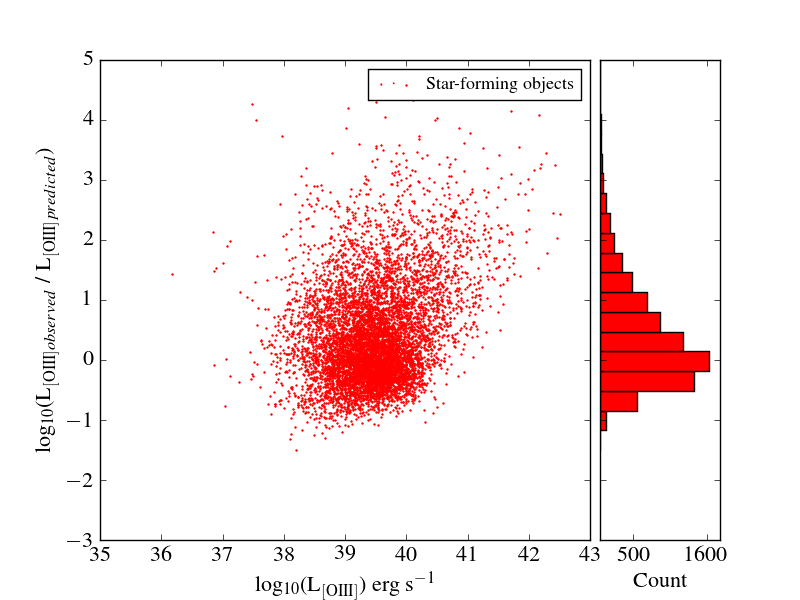}}
        \resizebox{0.497\hsize}{!}{\includegraphics{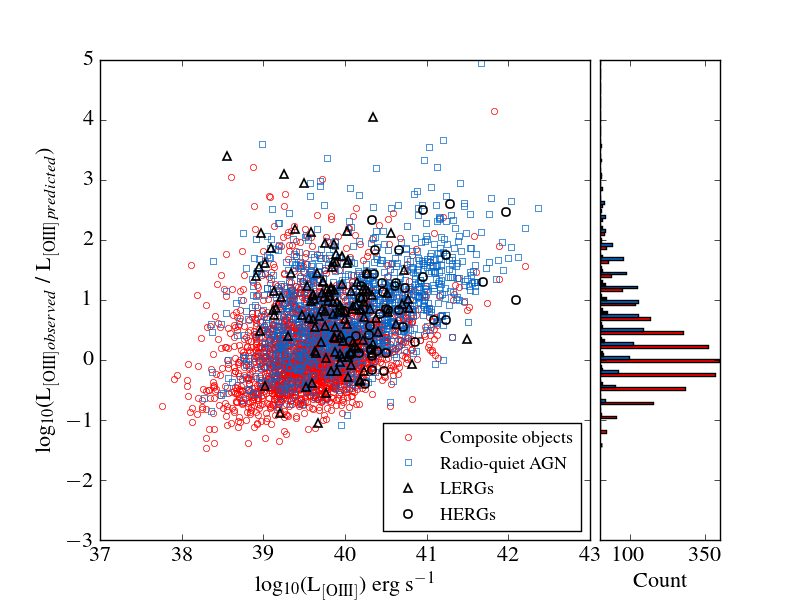}}
        \caption{The top graph shows the relation between L$_{\mathrm{[OIII]}}$ and $L_{250}$ for star-forming objects and our sample sources that are significantly (3$\sigma$) detected in their [OIII] emission-lines. In the bottom the distribution of the residual luminosities of the star-forming objects (left) and detected AGN in our sample (right) as a function of $L_{\mathrm{[OIII]}}$ together with a histogram of the counts are seen. The dashed line on the top panel shows the best fit for star forming objects. Star-forming objects are plotted as red crosses, radio-quiet AGN as blue squares, HERGs as open black circles and LERGs as open black triangles. }
\label{residual2}
\end{center}\vspace{-1em}
\end{figure*}

\begin{figure*}
\begin{center}
\scalebox{1}{
\begin{tabular}{c}
\centerline{\hspace{-0.9em}\includegraphics[width=15cm,height=15cm,angle=0,keepaspectratio]{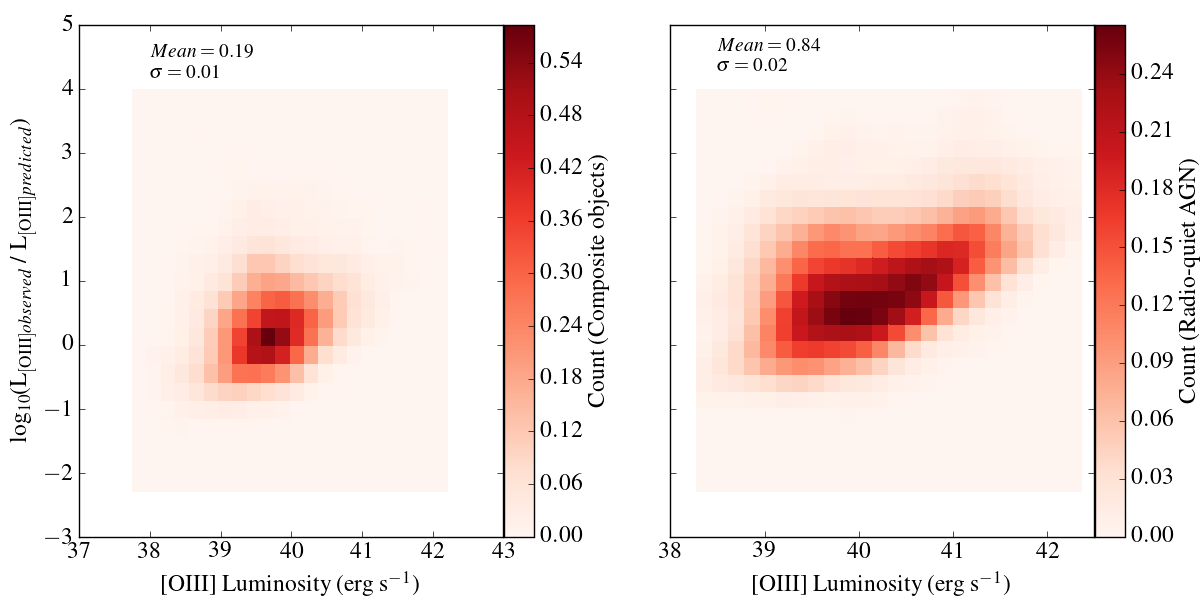}}\\
\end{tabular}}
\caption[.]{The comparison of density maps of the residuals ($log_{10}$(L$_{[\mathrm{OIII}]\,\, observed}$ / L$_{[\mathrm{OIII}]\,\, measured}$)) both for the composites (left) and radio-quiet AGN (right) is shown. The mean values of these calculated for both samples are also shown together with the errors on them derived using the bootstrap technique. \label{density_map} }
\end{center}
\end{figure*}

In the same way this relation was used to derive $L_{\mathrm{[OIII]}}$ for all objects in the AGN sample and for the star-forming objects. In order to understand the effect of star formation on the [OIII] luminosities we subtracted these estimates from the actual $L_{\mathrm{[OIII]}}$ measurements. In Figure \ref{residual2} the results of this analysis are shown. In the top panel we see that there is a correlation between log$_{10}$($L_{\mathrm{[OIII]}}$) and log$_{10}$($L_{250}$). However, radio-quiet and some radio-loud objects clearly have higher emission-line luminosities in comparison with star-forming sources. In the bottom plots, a boost in log$_{10}$($L_{\mathrm{[OIII]}}$) of the sources stands out when we compare with the distribution of star-forming objects seen in the left panel, although we see some trend of accumulation of sources around a residual of 0. To determine the cause of this the density maps for composite objects and radio-quiet AGN were produced. We also calculated the mean value of the residuals for both samples together with the error in these using the bootstrap method, which all are shown in Fig \ref{density_map}. It is clear in this figure that the composites peak around zero value of the residuals (the mean of residuals is 0.19 with $\sigma$=0.01) whereas the radio-quiet AGN are distributed above zero (the mean of residuals is 0.84 with $\sigma$=0.02). 

These results shows that 22-$\mu$m luminosity cannot be used as an AGN power estimator for this sample, \citep[a similar result was recently reported by][]{2013ref71} but that [OIII] emission is not dominated by star formation. Therefore, we used [OIII] luminosity in the calculation of the AGN powers. In order to compute radiative power we used the following relation: P$_{\mathrm{AGN}}$ = 3500$\times$ $L_{\mathrm{[OIII]}}$ given by \cite{2004ref5} which takes into account an average value for dust extinction. We can see in both Fig. \ref{residual1} and \ref{residual2} bottom left panels that there is a scatter to positive residual values between $L_{250}$ and $L_{\mathrm{22}}$, and $L_{\mathrm{[OIII]}}$. At 22 $\mu$m star formation dominates the band. Conversely, [OIII] is predominantly due to AGN with only a small contribution from SF.

To derive the jet powers, we utilised the relation $Q$(W) = 3$\times$10$^{38}$$\times f$$L_{151}^{6/7}$, where $L_{151}$ is the luminosity at 151 MHz, in units of $10^{28}$W Hz$^{-1}$ sr$^{-1}$, given by \cite{1999ref75} and $f$ is 10$^{3/2}$ \citep[see][]{2007ref177}. We derived the fluxes at 151 MHz using available fluxes at 1.4-GHz assuming the spectral index\footnote{In order to be consistent we use the same spectral index for all galaxies in the sample. However, we also used the GMRT estimates where available as a sanity check and found that this gave the same results.} to be 0.8. The jet power provides the best estimate of the AGN power for LERGs. It is worth noting that in principle the environment should be taken into account carefully in the conversion between radio luminosity and jet power \citep[e.g.][]{2014ref89}. However, we cannot do this with the data available to us. For HERGs we combined the calculated jet powers and radiative powers to obtain the AGN power as discussed by \cite{2014ref50}. AGN powers used in this work are summarised in Table \ref{agn-table}.

\begin{table}
\begin{tabular}{ccc}
\hline
Population type&&AGN Power\\
\hline
Radio-loud AGN&LERGs&Jet power\\
&HERGs&Jet power+Radiative power\\
Radio-quiet AGN&&Radiative power\\
\hline
\end{tabular}
\caption[]{Table shows AGN power definitions for different type of objects used in this work. As previously explained, Jet power was derived using extrapolated 151 MHz flux densities of both HERGs and LERGs. Radiative power was derived using [OIII] emission line luminosities for radio-quiet AGN sample and HERGs. }{\label{agn-table}}
\end{table}

\cite{1992ref170}, \cite{2014ref171} and \cite{2015ref172} have found that for a given AGN continuum luminosity radio-loud AGN have higher [OIII] magnitudes or fluxes than radio-quiet AGN as a result of having different Eddington accretion rates. Most of our radio-loud AGN contain LERGs which have different accretion mode than typical AGN. With regard to the SFR-AGN relation discussed in the following section HERGs only dominate the third and fourth AGN power bins. If we take into account the bias mentioned above we would expect HERGs to be included in AGN power bins lower than their current bins. This would not make much difference to our results since they would still have lower SFRs than their radio-quiet counterparts considering the trend of their SFRs as a function of AGN power.

%%%%%%%%%%%%%%%%%%%%%%%%%%%%%%%%%%%%%%%%%%%%%%%%%%%%%%%%%%%%%%%%%%%%%%%%%%%%%%%%%%%%%

\section{Results}
\subsection{Stacking SFR: Radio-loud and radio-quiet comparison}
\begin{figure*}
\begin{center}
\scalebox{1}{
\begin{tabular}{c}
%\centerline{\hspace{-0.9em}\includegraphics[width=17cm,height=17cm,angle=0,keepaspectratio]{MASS-SFR-ALL.png}}\\
\centerline{\hspace{-0.9em}\includegraphics[width=17cm,height=17cm,angle=0,keepaspectratio]{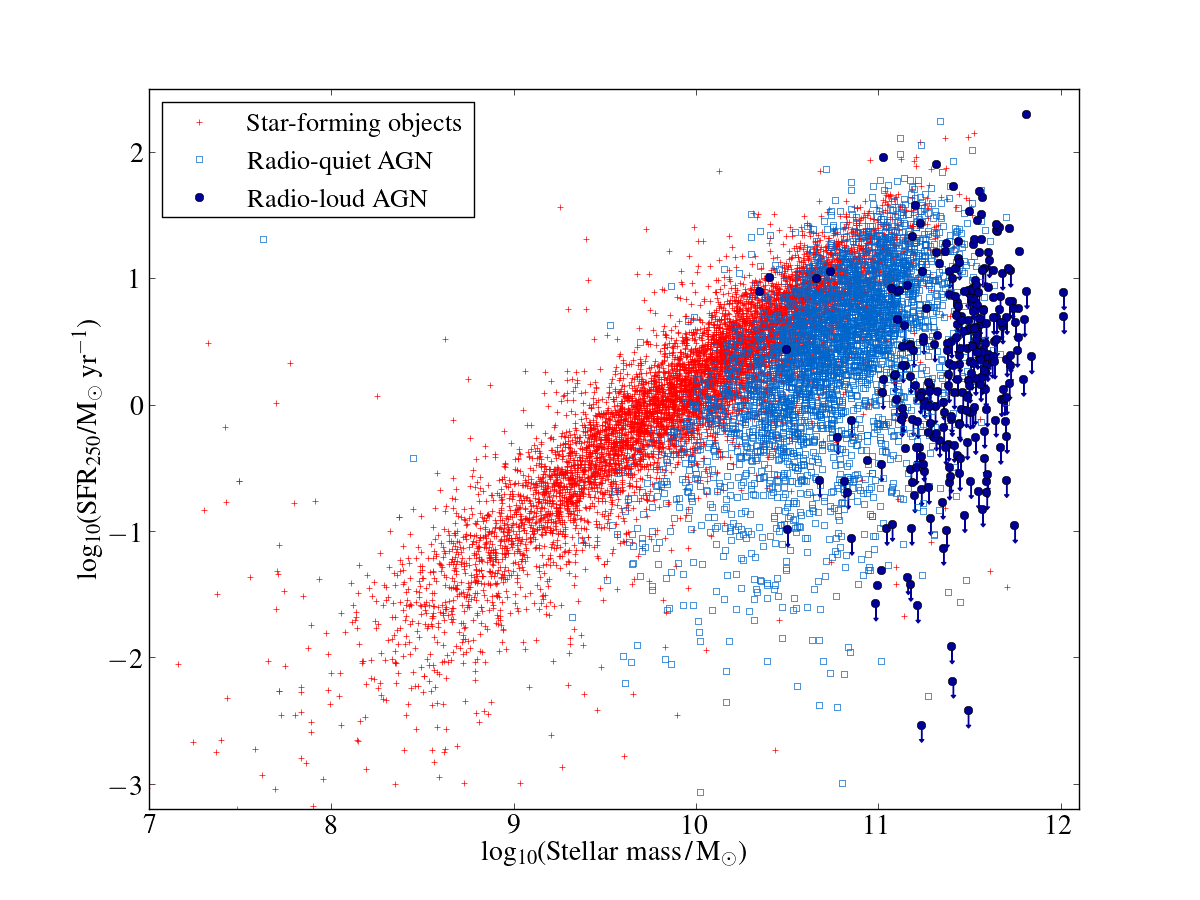}}\\
\end{tabular} }
\caption[]{The distributions of SFRs of the sources in our sample detected in 250 $\mu$m $\textit{Herschel}$ band and star-forming galaxies plotted as a function of their stellar masses. Star-forming objects are plotted as red crosses, radio-quiet AGN as blue squares and radio-loud AGN as navy filled circles. Radio-loud galaxies not detected at 5$\sigma$ the $\textit{Herschel}$ 250-$\mu$m band are indicated with arrows. We do not show these for radio-quiet objects in order to present the distribution of the sources clearly as the number of radio-quiet AGN is high.}
\label{mass-sfr}
\end{center}
\end{figure*}

In Figure \ref{mass-sfr} we show the distribution of SFRs of the star-forming objects, and radio-loud and radio-quiet AGN sample against their stellar masses. From the figure it can also be seen that radio-quiet and radio-loud AGN have lower SFRs compared to star-forming objects. Radio-loud galaxies not detected at 5$\sigma$ in $\textit{Herschel}$ 250-$\mu$m band are indicated with arrows. 

To compare the star formation properties of radio-loud and radio-quiet AGN as a function of AGN power we used the $\textit{Herschel}$ 250-$\mu$m band fluxes. However, the majority of the sources are not detected at the 5$\sigma$ level in the $\textit{Herschel}$ 250-$\mu$m band. Therefore, we stacked SFRs, SSFRs and stellar masses of the sources in 5 AGN power bins. 

We determined the mean values of the SFRs and SSFRs of the samples in all bins. While calculating SFRs we first found the mean value of $L_{250}$ in each bin and then calculated the mean SFR in the corresponding bin by using the relationship derived in Section 2.2. This allows us to not be biased against sources that are weak or not formally detected. For the derivation of SSFRs, we divided the sum of $L_{250}$ by the sum of the stellar mass in individual bins. This gave us the mean SSFR in the corresponding bins. The errors on these quantities were calculated using the bootstrap technique which enables us to derive the errors empirically, making no assumptions about the distribution of the luminosity. The 68 percentile confidence intervals derived from the bootstrap samples were used as errors on SFRs. Errors on SSFRs were calculated by combining the uncertainties throughout the bootstrap analysis for SFRs and stellar masses in quadrature. All these estimates for both radio-loud and radio-quiet samples can be found in Table \ref{tbl1}.

\begin{figure}
\begin{center}
        \resizebox{1.05\hsize}{!}{\includegraphics{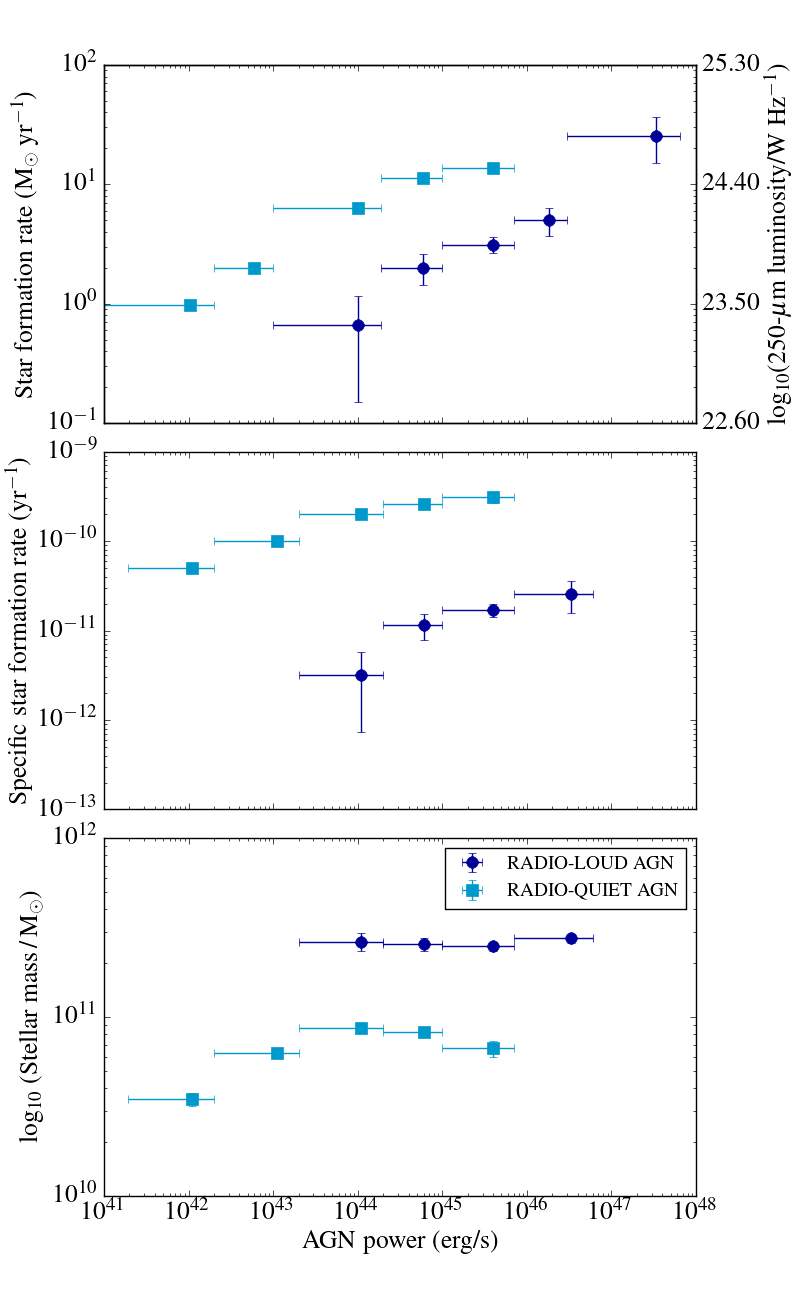}}\\
        %\resizebox{1.05\hsize}{!}{\includegraphics{MASS-PWR-L-Q.png}}\\ 
        %\resizebox{1.05\hsize}{!}{\includegraphics{SSFR-PWR-L-Q-initial.png}}\\
       %\resizebox{0.497\hsize}{!}{\includegraphics{}}
        \caption{The results of stacking analysis of radio-loud and radio-quiet AGN. From bottom to top the graphs show: the comparison of SSFRs, stellar masses and SFRs of the samples stacked in their AGN powers. Since we included sources that do not have mass measurements in the stacking analysis of SFRs, this allowed us to have more sources and accordingly more bins for radio-loud AGN. The first two AGN power bins did not have enough radio-loud AGN but we still show the mean value of SFR corresponding to radio-quiet AGN.}
\label{l-q1}
\end{center}\vspace{-1em}
\end{figure}

It can clearly be seen in Figure \ref{l-q1} that SFRs and SSFRs increase with increasing AGN power for both radio-loud and radio-quiet samples. We also show the mean stellar masses for the bins to see the effect of mass on the observed relation between SFR and AGN power. Most of our sources in the radio-loud AGN sample are LERGs hosted by galaxies that are presumably massive ellipticals as indicated in (Fig. \ref{l-q1}, \ref{l-q2} and \ref{l-h}). We see that radio-loud AGN reside in more massive galaxies in comparison with their radio-quiet counterparts. Our results show that in general radio-quiet AGN of a given AGN power are found to have higher SFR and SSFRs than radio-loud AGN. The difference between the SSFRs of radio-loud and radio-quiet AGN goes up to an order of magnitude. 
%\rew{In the last bin where we have sources with high luminosities, the two samples have similar values of SFRs with radio-loud AGN showing slightly higher values. This may be due to HERGs dominating this bin.}

Stacked measurements in confused images can be biased by the presence of correlated sources because the large PSF can include flux from nearby sources. Several methods have been proposed to account for this bias, including the flux measurements in GAMA apertures by \cite{2012ref168}. This method explicitly deblends confused sources and divides the blended flux between them using PSF information, so that average flux is conserved and not counted multiple times in the stack. We checked for the effects of clustering in our stacks by comparing to the results obtained from average flux measurements in the catalogue given by \cite{2012ref168}, for the 1758 objects overlapping between the two samples. Average fluxes for the matched sample are slightly higher using the deblended apertures method compared with our stacking method, which indicates that our method is not biased by the effects of clustering and the results are robust. The reason for the slightly higher fluxes using the catalogue from Bourne et al. is likely because they account for extended flux outside of the central beam. These results are consistent with the trends in our full stacked samples.

\subsubsection{Effects of the intrinsic correlation between SFR and redshift, and the Malmquist bias} 
\begin{figure}
\begin{center}
\scalebox{1.0}{
\begin{tabular}{c}
\centerline{\hspace{-0.7em}\includegraphics[width=10.5cm,height=10.5cm,angle=0,keepaspectratio]{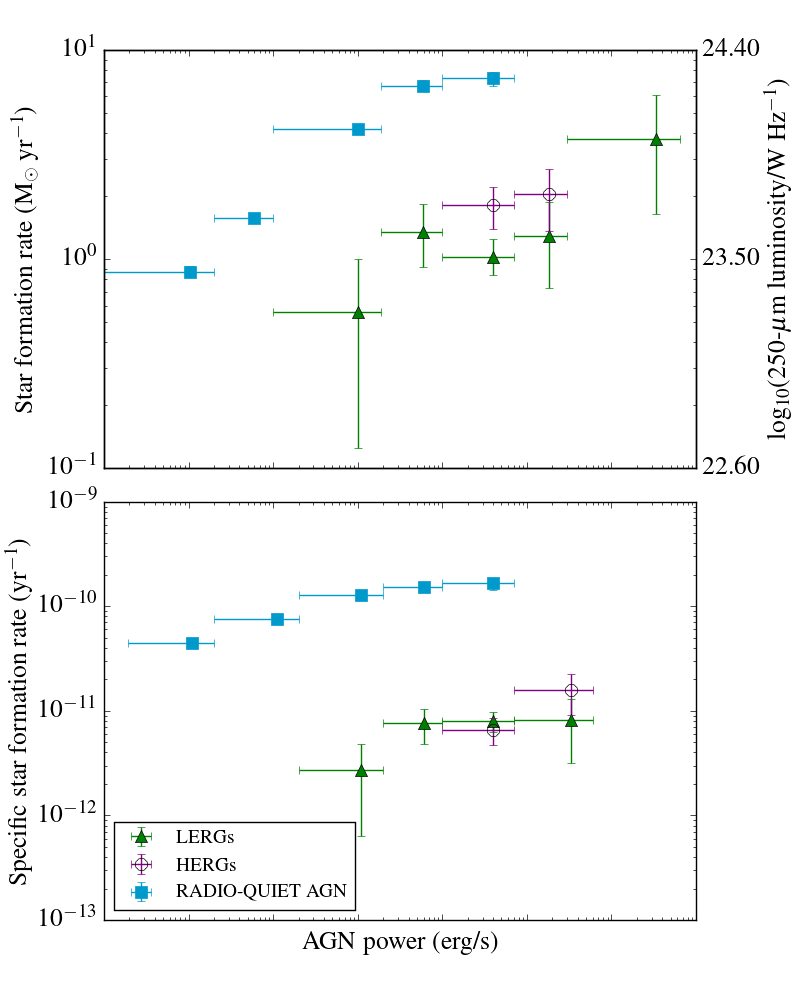}}\\
%\centerline{\hspace{-0.9em}\includegraphics[width=10.5cm,height=10.5cm,angle=0,keepaspectratio]{SSFR-PWR-L-Q.png}}\\
\end{tabular}}
\caption[.]{The results of stacking analysis of radio-loud and radio-quiet AGN where the intrinsic relation between $z$ and SSFR given by \cite{2010ref142} was taken into account. \label{z-effect} }
\end{center}
\end{figure}

It has been shown that SSFRs of galaxies evolve with redshift \citep[e.g.][and references therein]{2010ref142}. The effect of this intrinsic correlation found between SSFR and $z$ should be taken into account. In order to check if our results are affected by this relation we tried deriving the relation between SSFRs and $z$ using our star-forming objects but found an unrealistically strong correlation between SFR and $z$. The reason for this is that the sample used here is biased in the sense of having star-forming objects that are selected using optical emission lines. This causes us to only select objects with high star formation at high redshifts. We are not able to match our samples in redshift and stellar mass because of the small size of the radio-loud AGN sample. Instead we used the relation given by \cite{2010ref142} where they showed that late-type galaxies have SSFR $\propto$ (1+$z$)$^{3.36}$. We then took this relation into account and re-derived the relationships of SFRs and SSFRs as a function of AGN power (Fig. \ref{z-effect}). Comparison of this with our initial results (Fig. \ref{l-q1}) clearly shows that our results are not affected by the intrinsic correlation between $z$ and SSFR because the trends we initially observed between SFR/SSFR and AGN power do not change. Although we assume the same SSFR evolution with redshift for all types of galaxies here, radio-loud AGN have been found in more evolved galaxies in comparison to radio-quiet AGN and star-forming objects in the local Universe. Accordingly a less strong evolution is expected for these sources. Therefore, by assuming a stronger correlation between SSFR and $z$ for radio-loud AGN we may overestimate this effect on these particular galaxies.

We note that Malmquist bias does not affect our derived SFRs in the stacking analysis since we include all objects, including non-detections, in the stacking. The parent sample is of course flux-limited in [OIII] and/or radio luminosity, giving a correlation between AGN power and redshift, but this is included by taking the SSFR$-z$ relation into account described above.

\begin{figure}
\begin{center}
\scalebox{0.85}{
\begin{tabular}{c}
\centerline{\hspace{-0.9em}\includegraphics[width=11cm,height=11cm,angle=0,keepaspectratio]{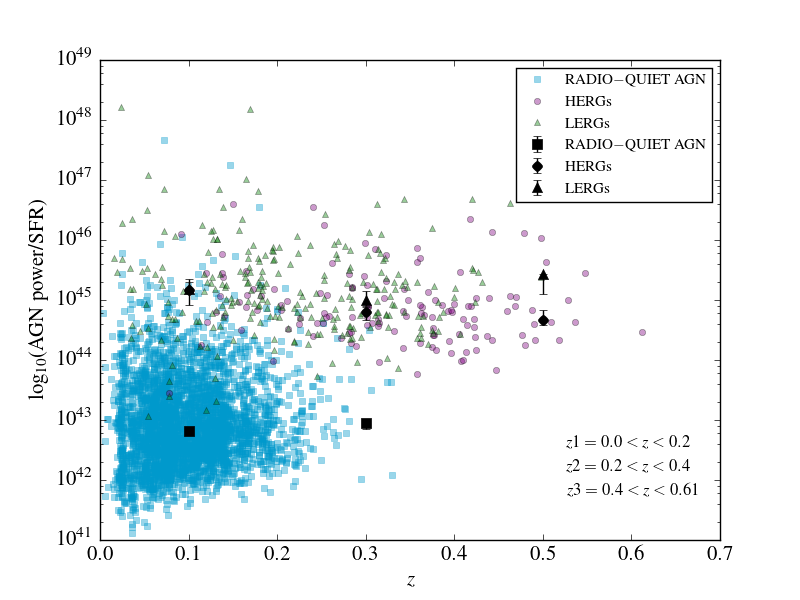}}\\
\end{tabular}}
\caption[.]{The distribution of AGN power/SFR versus redshift for radio-quiet AGN, HERGs and LERGs. Black points show the median estimations of the AGN power/SFR ratio for all population for 3 redshift bins. Errors on these estimations were derived from the bootstrapping. \label{agn-sfr-z}}
\end{center}
\end{figure}

As indicated above due to limited sample size we are not able to carry out the stacking analysis by binning in both redshift and AGN power. However, we implement a separate analysis in order to check the effect of universal SFR/SSFR evolution with redshift.  For this we estimated AGN power/SFR ratios for HERGs, LERGs and radio-quiet AGN. In Figure \ref{agn-sfr-z} we show the distribution of the AGN power/SFR ratio for all populations in our sample. Due to a large scatter, we binned the data in 3 redshift bins and calculated the median of the AGN power/SFR ratio for each bin for all populations. Errors on these values were derived by the bootstrapping. In the first 2 redshift bins radio-quiet AGN have lower AGN power/SFR ratio than HERGs and LERGs. This indicates that at all redshifts radio-quiet AGN have higher SFR compared to HERGs and LERGs. Somewhat surprisingly we find that HERGs and LERGs have similar AGN power/SFR ratios in most of the redshift bins. Therefore, we conclude that the difference between SFR of radio-loud and radio-quiet AGN is not due to the redshift effect.

\subsubsection{Effect of the composite objects}
\begin{figure}
\begin{center}
        \resizebox{1.05\hsize}{!}{\includegraphics{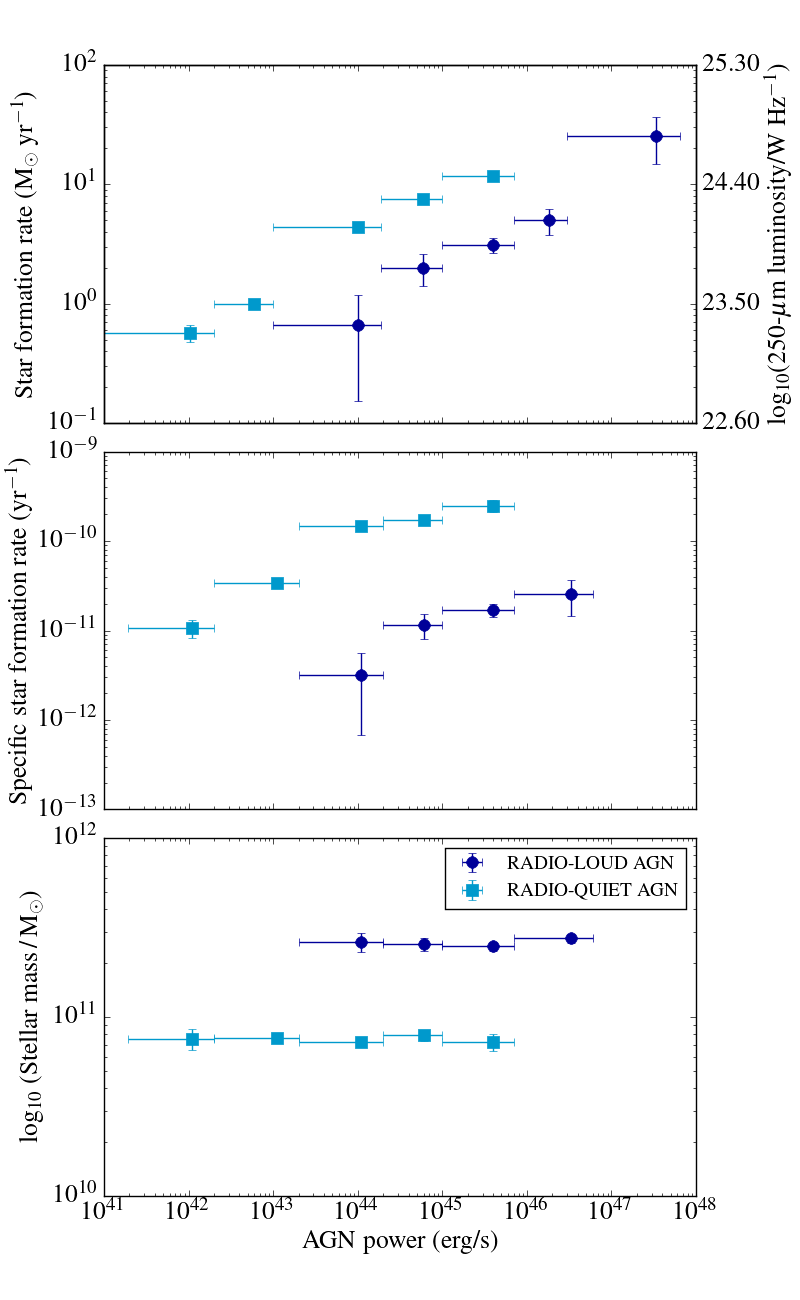}}\\
        %\resizebox{1.05\hsize}{!}{\includegraphics{MASS-PWR-L-Q-NO-COMPOSITE.png}}\\
        %\resizebox{1.05\hsize}{!}{\includegraphics{SSFR-PWR-L-Q-NO-COMPOSITE.png}}\\
        \caption{The result of stacking analysis of radio-loud and -quiet AGN where composite objects are not included. Similar to Figure \ref{l-q1} from bottom to top the graphs show: the comparison of SSFRs, stellar masses and SFRs of the samples stacked in their AGN powers.}
\label{l-q2}
\end{center}\vspace{-1em}
\end{figure}

Composites are sources that have emission-line characteristics intermediate between pure AGN and star-forming objects, have intermediate mass and SFR and lie on a typical Baldwin-Phillips-Terlevich diagram \citep[BPT; ][]{1981ref115} between pure star-forming objects and AGN \citep[e.g.][]{2007ref110,2011ref111}. We implemented the stacking process described above excluding the composite objects from the sample to see if our results were affected by the inclusion of these objects.  A striking difference would mean that while selecting the composites we mostly pick sources with high levels of star formation. However, this comparison showed that there is no difference between the SFRs and SSFRs of the samples with and without the composite sources compare Figs. \ref{l-q1} and \ref{l-q2}). Furthermore,  \cite{2011ref70} demonstrated that the expected SSFR of star-forming objects is around 2.5$\times$10$^{-10}$ yr$^{-1}$. In both Figures \ref{l-q1} and \ref{l-q2} the highest mean SSFRs are lower than this value. This also tells us that the composites are not dominated by star forming galaxies. Therefore, we kept the composites in the radio-quiet AGN sample for all analyses presented in this paper. Corresponding measurements are given in Table \ref{tbl2}.

\begin{table*}
\begin{tabular}{cccccccccc}
\hline
Sample & AGN power bins & Mean $z$ & N & Mean SFR & AGN power bins & Mean $z$ & N & Mean SSFR & Mean stellar mass \\        
&erg s$^{-1}$ &&&M$_{\odot}$ yr$^{-1}$&erg s$^{-1}$&&&($\times10^{-11}$ yr$^{-1}$)&($\times10^{+11}$ M$_{\odot}$)\\
\hline
Radio-quiet AGN& 1.0e+41$-$2.0e+42 & 0.03 & 170 & 0.98 $_{(0.06)}^{(0.06)}$& 1.9e+41$-$2.0e+42 & 0.03 & 170 & 4.99 $_{(0.50)}^{(0.50)}$ & 0.35 $_{(0.03)}^{(0.03)}$\\
 & 2.0e+42$-$1.0e+43 & 0.06 & 1099 & 1.99 $_{(0.05)}^{(0.05)}$&  2.0e+42$-$2.0e+43 & 0.08 & 1864 & 9.97 $_{(0.26)}^{(0.28)}$ & 0.63 $_{(0.01)}^{(0.01)}$\\
 & 1.0e+43$-$1.9e+44 & 0.11 & 2082 & 6.29 $_{(0.11)}^{(0.10)}$  & 2.0e+43$-$2.0e+44 & 0.13 & 1333 & 20.15 $_{(0.59)}^{(0.62)}$ & 0.87 $_{(0.02)}^{(0.02)}$\\
 & 1.9e+44$-$9.8e+44 & 0.14 & 278 & 11.37 $_{(0.70)}^{(0.69)}$ & 2.0e+44$-$1.0e+45 & 0.15 & 260 & 25.69 $_{(2.08)}^{(2.09)}$ & 0.82 $_{(0.04)}^{(0.04)}$ \\
 & 9.8e+44$-$7.0e+45 & 0.18 & 50 & 13.66 $_{(1.15)}^{(1.12)}$& 1.0e+45$-$7.0e+45 & 0.18 & 49 & 30.86 $_{(4.24)}^{(3.99)}$ & 0.67 $_{(0.08)}^{(0.07)}$ \\
Radio-loud AGN & 1.0e+43$-$1.9e+44 & 0.04 & 15 & 0.66 $_{(0.51)}^{(0.51)}$& 2.0e+43$-$2.0e+44 & 0.05 & 16 & 0.32 $_{(0.25)}^{(0.25)}$ & 2.64 $_{(0.29)}^{(0.29)}$ \\
 & 1.9e+44$-$9.8e+44 & 0.10 & 57 & 2.01 $_{(0.58)}^{(0.62)}$& 2.0e+44$-$1.0e+45 & 0.11 & 60 & 1.16 $_{(0.38)}^{(0.36)}$ & 2.56 $_{(0.21)}^{(0.21)}$ \\
 & 9.8e+44$-$7.0e+45 & 0.25 & 342 & 3.12 $_{(0.46)}^{(0.47)}$& 1.0e+45$-$7.0e+45 & 0.22 & 266 & 1.70 $_{(0.31)}^{(0.27)}$ & 2.48 $_{(0.14)}^{(0.13)}$ \\
 & 7.0e+45$-$3.0e+46 & 0.34 & 178 & 5.05 $_{(1.23)}^{(1.27)}$& 7.0e+45$-$6.0e+46 & 0.26 & 93 & 2.59 $_{(1.05)}^{(1.09)}$ & 2.75 $_{(0.18)}^{(0.16)}$ \\
 & 3.0e+46$-$6.5e+47 & 0.35 & 20 & 25.38 $_{(10.44)}^{(9.64)}$&$-$&$-$&$-$&$-$&$-$\\
HERGs & 9.8e+44$-$7.0e+45 & 0.30 & 120 & 4.85 $_{(1.07)}^{(1.02)}$&  1.0e+45$-$7.0e+45 & 0.23    &68 & 1.43 $_{(0.42)}^{(0.40)}$ & 3.05 $_{(0.44)}^{(0.41)}$ \\   
      & 7.0e+45$-$3.0e+46 & 0.40 & 75 & 7.27 $_{(2.11)}^{(2.45)}$&     7.0e+45$-$6.0e+46 & 0.27  &22 & 3.88 $_{(1.65)}^{(1.76)}$ & 2.91 $_{(0.38)}^{(0.36)}$ \\   
LERGs & 1.0e+43$-$1.9e+44 & 0.05 & 15 & 0.66 $_{(0.51)}^{(0.51)}$&   2.0e+43$-$2.0e+44 & 0.05    &16 & 0.32 $_{(0.25)}^{(0.25)}$ & 2.64 $_{(0.32)}^{(0.31)}$ \\   
      & 1.9e+44$-$9.8e+44 & 0.10 & 51 & 1.94 $_{(0.70)}^{(0.66)}$& 2.0e+44$-$1.0e+45 & 0.10      &53 & 1.10 $_{(0.36)}^{(0.38)}$ & 2.56 $_{(0.23)}^{(0.22)}$ \\   
      & 9.8e+44$-$7.0e+45 & 0.23 & 222 & 2.23 $_{(0.41)}^{(0.50)}$ & 1.0e+45$-$7.0e+45 & 0.21   &198 & 1.67 $_{(0.36)}^{(0.32)}$ & 2.29 $_{(0.12)}^{(0.11)}$ \\  
      & 7.0e+45$-$3.0e+46 & 0.30 & 103 & 3.49 $_{(1.49)}^{(1.59)}$&    7.0e+45$-$6.0e+46 &0.25   &71 & 1.91 $_{(1.29)}^{(1.24)}$ & 2.70 $_{(0.19)}^{(0.18)}$ \\   
      & 3.0e+46$-$6.5e+47 & 0.35 & 13 & 11.65 $_{(7.11)}^{(7.23)}$&$-$&$-$&$-$&$-$&$-$\\  
\hline                                                                                         
\end{tabular}
\caption[]{The results of the stacking analysis regarding our samples. In column 2 the chosen AGN power bins are shown. $N$ indicates number of sources included in each bin. The mean measurements of SFRs in each bin with their errors calculated by the bootstrap technique are presented in column 5. New AGN power bins were defined for the mean measurements of SSFRs and stellar mass as the number of sources in the sample decreased by excluding sources that do not have stellar mass measurements. In column 6 the new AGN power bins are shown. The mean SSFRs and the mean stellar mass for each bin with their errors are shown in column 9 and 10, respectively.}{\label{tbl1}}
\end{table*}

\begin{table*}
\begin{tabular}{cccccccccc}
\hline
Sample & AGN power bins & Mean $z$ & N & Mean SFR& AGN power bins & Mean $z$ & N & Mean SSFR & Mean stellar mass \\             
&erg s$^{-1}$ &&&M$_{\odot}$ yr$^{-1}$&erg s$^{-1}$&&&($\times10^{-11}$ yr$^{-1}$)&($\times10^{+11}$ M$_{\odot}$)\\
\hline
Radio-quiet AGN & 1.0e+41$-$2.0e+42 & 0.02 & 23 & 0.57 $_{(0.10)}^{(0.09)}$&1.9e+41$-$2.0e+42 & 0.02 & 23 & 1.07 $_{(0.24)}^{(0.23)}$ & 0.76 $_{(0.10)}^{(0.10)}$ \\ \
 & 2.0e+42$-$1.0e+43 & 0.05 & 268 & 1.00 $_{(0.07)}^{(0.06)}$& 2.0e+42$-$2.0e+43 & 0.06 & 443 & 3.35 $_{(0.21)}^{(0.20)}$ & 0.76 $_{(0.03)}^{(0.03)}$ \\  
 & 1.0e+43$-$1.9e+44 & 0.10 & 657 & 4.35 $_{(0.19)}^{(0.18)}$& 2.0e+43$-$2.0e+44 & 0.11 & 490 & 14.72 $_{(0.86)}^{(0.86)}$ & 0.73 $_{(0.03)}^{(0.03)}$ \\ 
 & 1.9e+44$-$9.8e+44 & 0.14 & 201 & 7.48 $_{(0.56)}^{(0.56)}$& 2.0e+44$-$1.0e+45 & 0.14 & 191 & 17.06 $_{(1.74)}^{(1.75)}$ & 0.79 $_{(0.05)}^{(0.05)}$ \\ 
 & 9.8e+44$-$7.0e+45 & 0.17 & 40 & 11.80 $_{(1.07)}^{(0.96)}$& 1.0e+45$-$7.0e+45 & 0.17 & 40 & 24.65 $_{(3.42)}^{(3.75)}$ & 0.72 $_{(0.08)}^{(0.09)}$ \\  
Radio-loud AGN & 1.0e+43$-$1.9e+44 & 0.05 & 15 & 0.66 $_{(0.51)}^{(0.52)}$& 2.0e+43$-$2.0e+44 & 0.05 & 16 & 0.32 $_{(0.25)}^{(0.25)}$ & 2.64 $_{(0.33)}^{(0.32)}$ \\   
 & 1.9e+44$-$9.8e+44 & 0.10 & 57 & 2.01 $_{(0.61)}^{(0.57)}$&2.0e+44$-$1.0e+45 & 0.11 & 60 & 1.16 $_{(0.36)}^{(0.35)}$ & 2.56 $_{(0.21)}^{(0.21)}$ \\   
 & 9.8e+44$-$7.0e+45 & 0.25 & 342 & 3.12 $_{(0.45)}^{(0.46)}$&1.0e+45$-$7.0e+45 & 0.22 & 266 & 1.70 $_{(0.27)}^{(0.27)}$ & 2.48 $_{(0.15)}^{(0.14)}$ \\  
 & 7.0e+45$-$3.0e+46 & 0.34 & 178 & 5.05 $_{(1.29)}^{(1.19)}$&7.0e+45$-$6.0e+46 & 0.26 & 93 & 2.59 $_{(1.12)}^{(1.07)}$ & 2.75 $_{(0.18)}^{(0.16)}$ \\   
 & 3.0e+46$-$6.5e+47 & 0.36 & 20 & 25.38 $_{(10.59)}^{(11.10)}$&$-$&$-$&$-$&$-$&$-$\\
\hline                                                                                         
\end{tabular}
\caption[]{The results of the stacking analysis of the sample where the composite objects are not included. In column 2 chosen AGN power bins are presented. $N$ indicates number of sources included in each bin. The mean measurements of SFRs in each bin with their errors calculated by the bootstrap technique are shown in column 5. New AGN power bins were defined for the mean measurements of SSFRs and stellar mass as the number of sources in the sample decreased by excluding sources that do not have stellar mass measurements. In column 6 the new AGN power bins are shown. The mean SSFRs and the mean stellar mass for each bin with their errors are presented in column 9 and 10, respectively.}{\label{tbl2}}
\end{table*}

\subsubsection{Stacking SFR: Radio-loud and radio-quiet comparison using mass-matched samples}

\begin{figure}
\begin{center}
\scalebox{0.85}{
\begin{tabular}{c}
\centerline{\hspace{-0.9em}\includegraphics[width=10.5cm,height=10.5cm,angle=0,keepaspectratio]{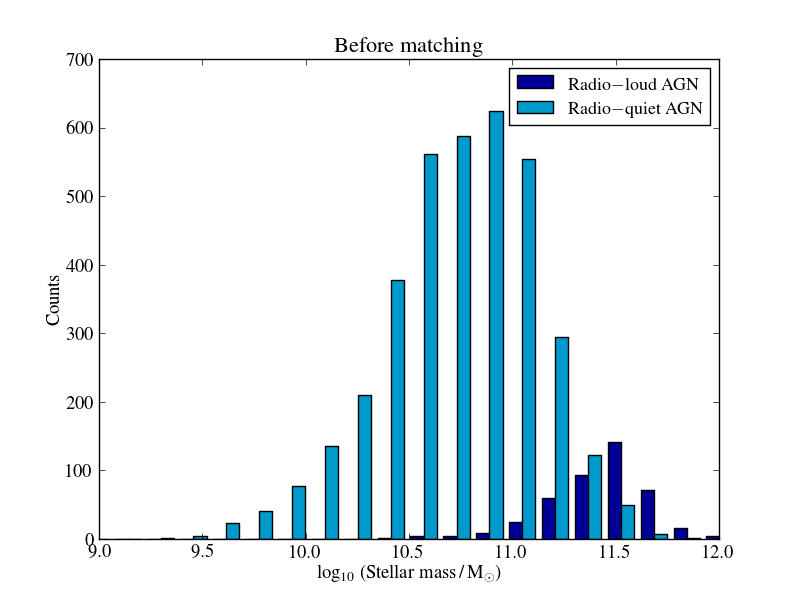}}\\
\centerline{\hspace{-0.9em}\includegraphics[width=10.5cm,height=10.5cm,angle=0,keepaspectratio]{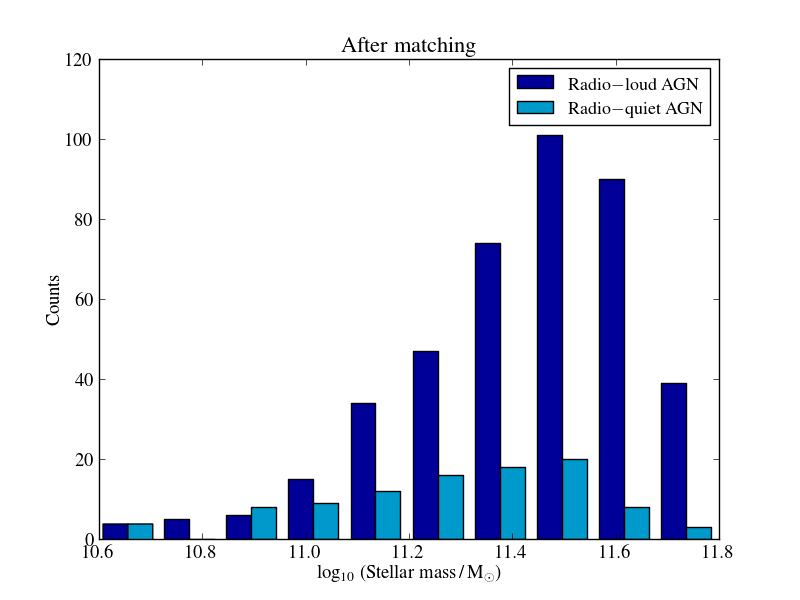}}\\
\end{tabular}}
\caption[.]{Stellar mass distribution of each AGN sample before (top plot) and after (bottom plot) the matching process. \label{matching-dist} }
\end{center}
\end{figure}

One of the possible effects that may cause the correlation between SFR and AGN power we see in our results is the mass of the host galaxy. In order to judge its influence and to eliminate it, we implemented the same analysis for sources matched in their stellar mass. We carried out mass matching using the method described by \cite{2013ref68}. In summary, we randomly discarded 1 per cent of the galaxies of the comparison sample (the radio-quiet AGN sample). We then ran a K-S test to compare the distributions of stellar masses for the main sample (the radio-loud AGN sample) and comparison sample (the radio-quiet AGN sample). Repeating this process N times, where N is the number of comparison sources, enables us to select the best reduced catalogue. Once we obtain the best matching catalogue, 1 per cent of the remaining sources in the comparison sample is discarded and the process is repeated. We repeat this process until the null hypothesis probability returned by the K-S test exceeds 10 per cent. This mass-matching process was implemented for the sources in each AGN power bin separately. Although this method provides us the best matched catalogue, we lose the majority of the original comparison sample. The loss is around 96 per cent so that our final sample of radio-quiet AGN contains 98 objects. Figure \ref{matching-dist} shows the distribution of stellar mass for each sample before and after the matching process.

\begin{figure}
\begin{center}
        %\resizebox{0.6\hsize}{!}{\includegraphics{SFR-MASS-MATCHED.png}}\\
        \resizebox{1.05\hsize}{!}{\includegraphics{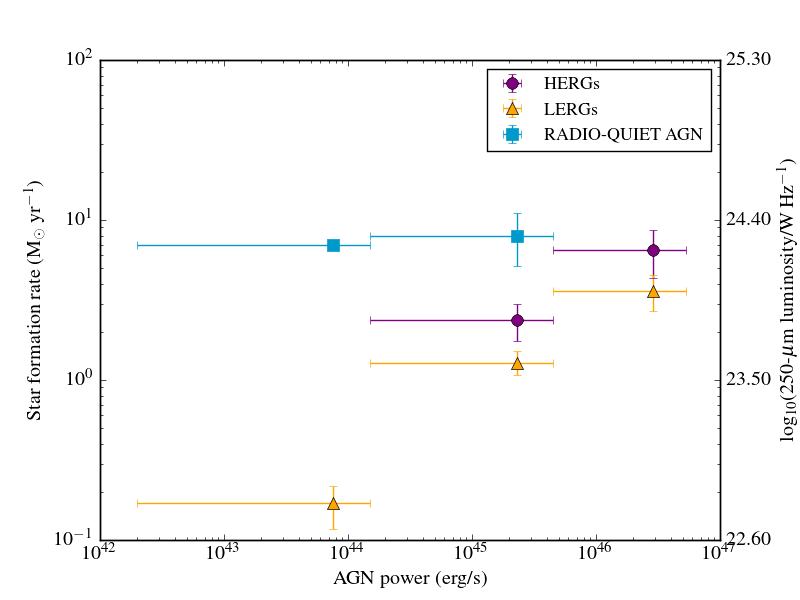}}\\
        %\resizebox{0.58\hsize}{!}{\includegraphics{SSFR-MASS-MATCHED.png}}\\
        \caption{The result of the SFR stacking analysis of radio-loud and radio-quiet AGN using mass-matched samples. The radio-loud objects are separated into emission-line class.}
\label{mass-matched}
\end{center}\vspace{-1em}
\end{figure}

The results of this analysis are presented in Figure \ref{mass-matched}. Although there seems to be a slight increase in SFR of radio-quiet AGN with AGN power the steepness of increase softened to a more gentle slope by leaving out the effect of galaxy mass. 
 
We also show emission-line radio galaxies separately in this figure. LERGs, in all bins, have the lowest rate of star formation as expected, since LERGs have been found in large ellipticals with little star formation \citep[e.g.][]{2005ref57}. For a given AGN power HERGs' host galaxies have comparatively higher star formation rates than LERGs but less than host galaxies of radio-quiet AGN. 

\begin{table*}
\begin{tabular}{ccccccc}
\hline
Sample&AGN power bins&Mean $z$&$N$&Mean SFR&Mean stellar mass\\
&erg s$^{-1}$&&& M$_{\odot}$ yr$^{-1}$&($\times10^{+11}$ M$_{\odot}$)\\
\hline
Radio-quiet AGN & 2.0e+42$-$1.5e+44 & 89 & 0.12 & 6.98 $_{(0.56)}^{(0.53)}$&2.08 $_{(0.10)}^{(0.11)}$ \\    
 & 1.5e+44$-$4.5e+45 & 8 & 0.18 & 7.93 $_{(3.01)}^{(3.10)}$& 3.25 $_{(0.49)}^{(0.51)}$ \\
HERGs & 1.5e+44$-$4.5e+45 & 59 & 0.21 & 2.38 $_{(0.70)}^{(0.61)}$&3.08 $_{(0.16)}^{(0.16)}$ \\
 & 4.5e+45$-$5.3e+46 & 36 & 0.27 & 6.45 $_{(2.14)}^{(1.70)}$& 3.28 $_{(0.17)}^{(0.19)}$ \\
LERGs & 2.0e+42$-$1.5e+44 & 7 & 0.04 & 0.17 $_{(0.05)}^{(0.05)}$& 1.12 $_{(0.12)}^{(0.11)}$ \\
 & 1.5e+44$-$4.5e+45 & 208 & 0.17 & 1.28 $_{(0.21)}^{(0.22)}$&2.63 $_{(0.08)}^{(0.08)}$ \\
 & 4.5e+45$-$5.3e+46 & 105 & 0.25 & 3.60 $_{(0.97)}^{(0.90)}$& 3.58 $_{(0.13)}^{(0.11)}$ \\
\hline
\end{tabular}
\caption[]{The stacking analysis results of radio-loud (HERGs and LERGs) and radio-quiet AGN samples matched in their stellar masses. Similar to Table \ref{tbl1} in column 2 chosen AGN power bins are shown. In column 3 the mean redshift measurements are given for each bin. $N$ indicates number of sources included in each bin. The mean measurements of SFRs and stellar mass in each bin with their errors calculated by the bootstrap technique are presented in column 5 and 6, respectively. We also show here emission-line radio galaxies included in each bin and their corresponding measurements as a reference.}{\label{mass-match-table}}
\end{table*}

\subsection{Correlation between black hole accretion rate and star formation rate}

If SFR and AGN power are coupled then a correlation should be expected between black hole accretion rates ($\dot{M}_{\textrm{BH}}$) and star formation rates ($\dot{M}_{\textrm{SFR}}$) of these samples. In order to assess this, $\dot{M}_{\textrm{BH}}$ were calculated using P$_{\textrm{AGN}}$=$\eta \dot{M}_{\textrm{BH}} c^2$ where $P_{\textrm{AGN}}$ is the AGN power, whose derivation was discussed in Section 2.4 {(see also Table \ref{agn-table})}, $\eta$ is an efficiency factor, $\dot{M}_{\textrm{BH}}$ is the black hole accretion rate and $c$ is the speed of light. We assumed the value of efficiency to be 0.1 for all sources. However, it should be noted that since the efficiency depends on the nature of accretion disk and accretion flows, a range of efficiencies can be found for different types of AGN. Especially in convection-dominated accretion flow \citep[CDAF; e.g.][]{2000ref123}, advection dominated accretion \citep[ADAF; e.g.][]{1995ref74} and adiabatic inflow-outflow solution \citep[ADIOS; e.g.][]{1999ref122} accretion flow models only a small fraction of the matter contributes to the mass accretion rate at the black hole because of turbulence and strong mass loss. For this reason, much lower radiative efficiencies ($\eta\ll0.1$) are expected but the efficiency of the jet-generation process is not known. On the contrary, thin-disk accretion onto a black hole may lead to high efficiency factors \citep[$\eta>0.1$,][]{1988ref101,2002ref124}. The use of a single efficiency factor is therefore only an approximation.

\begin{figure}
\begin{center}
\scalebox{0.85}{
\begin{tabular}{c}
\centerline{\hspace{-0.9em}\includegraphics[width=10.5cm,height=10.5cm,angle=0,keepaspectratio]{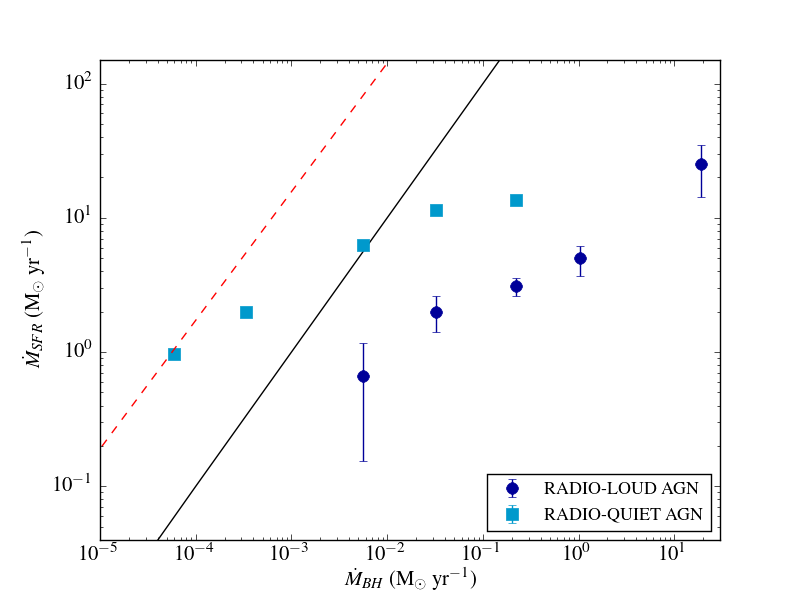}}\\
\end{tabular}}
\caption[.]{The relation between the rate of black hole growth and the rate of galaxy growth is presented for our AGN samples. Assuming $\dot{M}_{\textrm{SFR}}$ is proportional to bulge mass and $\dot{M}_{\textrm{BH}}$ is proportional to black hole mass, the black solid line represents the standard relation between the bulge and black hole mass. \citep[e.g.][]{2003ref18,2004ref5}. Low-power AGN have black holes growing slower than expected value and black holes in high-power AGN are growing faster. The red dashed line represents the relationship between $\dot{M}_{\textrm{SFR}}$ and $\dot{M}_{\textrm{BH}}$  that would be observed if star-forming objects were incorrectly classified as AGN. It can be seen that such contamination of the AGN sample cannot account for the observed trend. \label{accretion} }
\end{center}
\end{figure}

A proxy of the rate of black hole growth ($\dot{M}_{\textrm{BH}}$) as a function of star formation rate (a proxy of the rate of galaxy growth) is presented in Figure \ref{accretion} where we see a strong relationship between the rate of matter accreted onto black hole and that of star formation for both types of active galaxy samples. 

We show the standard relation between bulge and black hole mass \citep[e.g.][]{2003ref18,2004ref5} with black solid line for a comparison to the relations we see between the rate of black hole growth and galaxy growth for our samples. When we compare our results with the observed standard correlation suggests that low-power AGN have black holes growing slower than expected to maintain the M$_{\textrm{BH}}$$-$M$_{\textrm{Bulge}}$ relation (for example for a source with $\dot{M}_{\textrm{SFR}}$ = 9$\times$10$^{-1}$ M$_{\odot}$ yr$^{-1}$, the expected value of $\dot{M}_{\textrm{BH}}$ is around 10$^{-3}$ M$_{\odot}$ yr$^{-1}$) and black holes in high-power AGN are growing faster (for an object with $\dot{M}_{\textrm{SFR}}$ = 2.5$\times$10$^{1}$ M$_{\odot}$ yr$^{-1}$, the expected $\dot{M}_{\textrm{BH}}$ is to be about 9$\times$10$^{-1}$ M$_{\odot}$ yr$^{-1}$). The implications of this result are discussed in Section 4.1.

We also checked whether any possible contribution from star formation to the [OIII] luminosity can affect the relation that we obtained between SFR and $\dot{M}_{\textrm{BH}}$. The red dashed line shows the expected relationship between accretion rate, derived by using the [OIII] luminosity, and $\dot{M}_{\textrm{SFR}}$ for star-forming objects. The comparison of the red line with the results of stacking analysis for AGN demonstrates that the contribution from star formation to the [OIII] luminosity can only have an effect on sources in the lowest AGN power bin.

\begin{table}
\begin{tabular}{ccccccc}
\hline
&&Population type&$\dot{M}_{\textrm{BH}}$ range&SFR/$\dot{M}_{\textrm{BH}}$&&\\
\hline
&&Radio-quiet AGN&10$^{-5}$ $-$ 2$\times$10$^{-3}$         &15000-1500&&\\
             &&&2$\times$10$^{-3}$ $-$ 2$\times$10$^{-1}$&1500-100&&\\
&&Radio-loud AGN&10$^{-3}$ $-$ 2$\times$10$^{-1}$        &150-15&&\\
              &&&2$\times$10$^{-1}$ $-$ 2$\times$10$^{1}$&15-2.0&&\\
\hline
\end{tabular}
\caption[]{SFR/$\dot{M}_{\textrm{BH}}$ ratios for radio-loud and radio-quiet AGN are given for different range of accretion rates.}{\label{sfr/bhar}}
\end{table}

\subsection{Stacking AGN power: HERG and LERG comparison}

\begin{figure}
\begin{center}
        \resizebox{1.05\hsize}{!}{\includegraphics{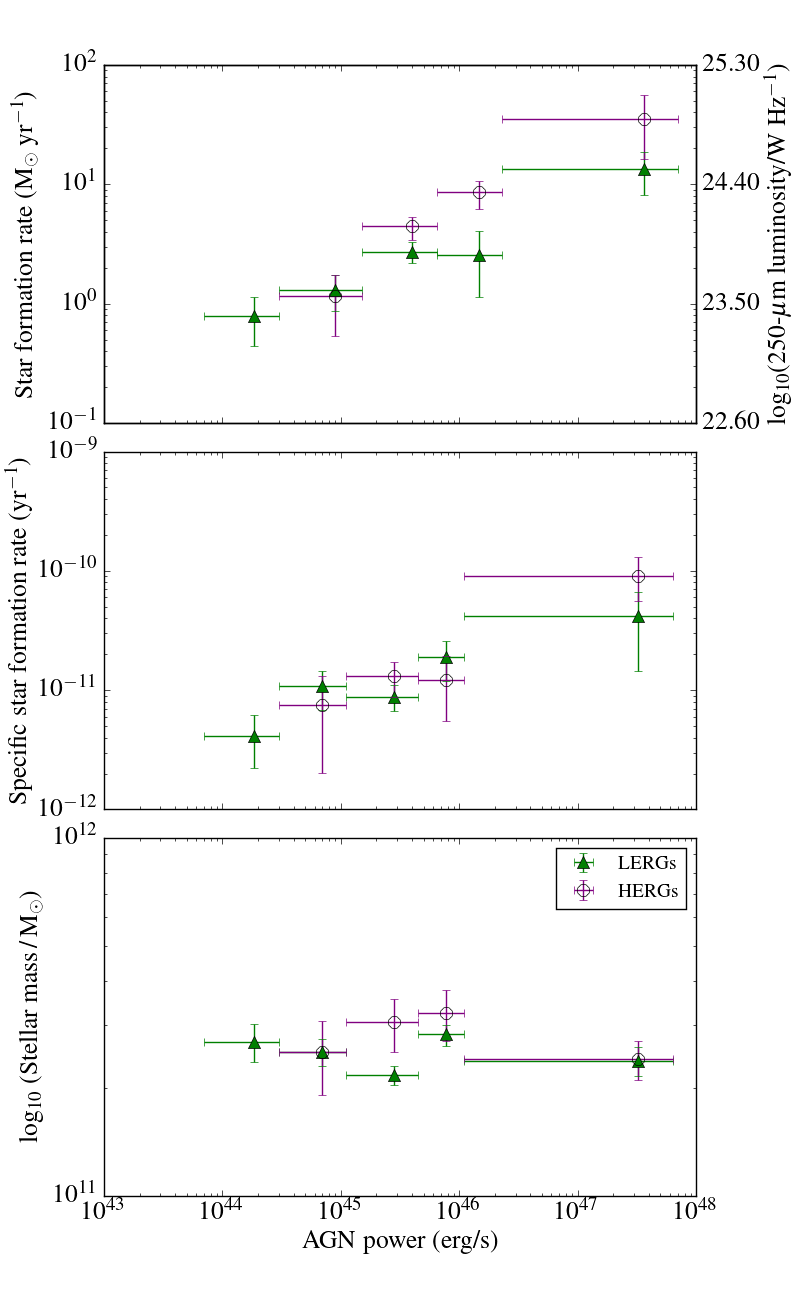}}\\
        %\resizebox{1.05\hsize}{!}{\includegraphics{MASS-PWR.png}}\\ 
        %\resizebox{1.05\hsize}{!}{\includegraphics{SSFR-PWR.png}}\\
        \caption{The result of stacking analysis of radio-loud AGN sample. Different AGN power bins were used in order to have the appropriate number of sources in each bin to be able to compare different emission classes. From bottom to top the graphs shows: the comparison of SSFRs, stellar masses and SFRs of the samples stacked in their AGN powers.}
\label{l-h}
\end{center}\vspace{-1em}
\end{figure}

The radio-loud AGN sample was investigated on its own to study the possible difference in the star formation properties of the sources as a function of emission-line type, discussed previously by \cite{2010ref7,2013ref9}. Since we want to compare the SFRs and SSFRs for HERGs and LERGs, we defined new AGN power bins to include the similar numbers of sources in each bin for both classes. We present the results in Figure \ref{l-h}. In the top graph we can see that the change in the SFRs is dependent on the AGN power. SFRs increase, for both HERGs and LERGs, with increasing AGN power. The increase in SSFRs for LERGs is more gentle in comparison to the increase of their SFRs whereas we still see a clear increasing trend of SSFRs for HERGs. In two out of three SFR bins (and three out of four bins of SSFR), LERGs have lower SFRs/SSFRs than HERGs, but the error bars are large.

Since the size of the radio-loud AGN sample is small we only have four AGN bins for the stacking analysis of SFR, and this is reduced to three bins for the estimates of SSFRs when the galaxies with no stellar mass estimates were excluded. Therefore, a quantitative analysis was carried out to find the magnitude of any possible difference between the SFRs of LERGs and HERGs. We divided the SFRs of LERGs by the SFRs of HERGs for each AGN power bin to calculate the SFR ratio and the corresponding errors on these ratios. The mean ratio was then computed using these estimates. The mean ratio of SFR$_{\mathrm{LERGs}}$/SFR$_{\mathrm{HERGs}}$ is 0.60$\pm$0.32. This result indicates that within the errors there is at most a slight difference between the SFRs of LERGs and HERGs of matched AGN power.

\begin{table*}
\begin{tabular}{cccccccccc}
\hline
Sample & AGN power bins & Mean $z$ & N & Mean SFR& AGN power bins & Mean $z$ & N & Mean SSFR & Mean stellar mass \\                         
&erg s$^{-1}$ &&&M$_{\odot}$ yr$^{-1}$&erg s$^{-1}$&&&($\times10^{-11}$ yr$^{-1}$)&($\times10^{+11}$ M$_{\odot}$)\\
\hline
LERGs & 7.0e+43$-$3e+44 & 0.05 & 22 & 0.78$_{0.35}^{0.35}$ & 7.0e+43$-$3.0e+44 & 0.05 & 22 & 0.41 $_{(0.19)}^{(0.21)}$ & 2.69 $_{(0.32)}^{(0.32)}$ \\    
 & 3.0e+44$-$1.5e+45 & 0.13 & 77 & 1.29$_{0.42}^{0.43}$& 3.0e+44$-$1.1e+45 & 0.12 & 54 & 1.09 $_{(0.41)}^{(0.37)}$ & 2.52 $_{(0.22)}^{(0.23)}$ \\  
 & 1.5e+45$-$6.5e+45 & 0.24 & 186 & 2.71$_{0.52}^{0.55}$& 1.1e+45$-$4.5e+45 & 0.21 & 149 & 0.87 $_{(0.19)}^{(0.24)}$ & 2.17 $_{(0.13)}^{(0.13)}$ \\
 & 6.5e+45$-$2.3e+46 & 0.30 & 101 & 2.57$_{1.44}^{1.46}$& 4.5e+45$-$1.1e+46 & 0.25 & 76 & 1.89 $_{(0.69)}^{(0.69)}$ & 2.83 $_{(0.21)}^{(0.18)}$ \\
 & 2.3e+46$-$7.0e+47 & 0.34 & 18 & 13.56$_{5.43}^{5.08}$& 1.1e+46$-$6.4e+47 & 0.25 & 37 & 4.14 $_{(2.71)}^{(2.54)}$ & 2.39 $_{(0.23)}^{(0.23)}$ \\
HERGs & 3.0e+44$-$1.5e+45 & 0.16 & 23 & 1.17$_{0.63}^{0.57}$&3.0e+44$-$1.1e+45 & 0.14 & 10 & 0.75 $_{(0.55)}^{(0.55)}$ & 2.52 $_{(0.61)}^{(0.57)}$ \\  
 & 1.5e+45$-$6.5e+45 & 0.32 & 96 & 4.43$_{1.03}^{0.94}$ & 1.1e+45$-$4.5e+45 & 0.22 & 51 & 1.30 $_{(0.42)}^{(0.41)}$ & 3.05 $_{(0.53)}^{(0.51)}$ \\
 & 6.5e+45$-$2.3e+46 & 0.39 & 78 & 8.62$_{2.42}^{2.12}$ & 4.5e+45$-$1.1e+46 & 0.27 & 26 & 1.23 $_{(0.68)}^{(0.71)}$ & 3.24 $_{(0.54)}^{(0.52)}$ \\
 & 2.3e+46$-$7.0e+47 & 0.40 & 11 & 35.25$_{19.06}^{21.24}$ & 1.1e+46$-$6.4e+47 & 0.28 & 10 & 9.09 $_{(3.55)}^{(3.85)}$ & 2.41 $_{(0.31)}^{(0.31)}$ \\
\hline
\end{tabular}
\caption[]{Table shows the stacking analysis results for HERGs and LERGs. In column 2 the chosen AGN power bins are shown. $N$ indicates number of sources included in each bin. The mean measurements of SFRs in each bin with their errors calculated by the bootstrap technique are presented in column 5. New AGN power bins were defined for the mean measurements of SSFRs and stellar mass as the number of sources in the sample decreased by excluding sources that do not have stellar mass measurements. In column 6 the new AGN power bins are shown. The mean SSFRs and the mean stellar mass for each bin with their errors are shown in column 9 and 10, respectively.}{\label{tbl2}}
\end{table*}

%%%%%%%%%%%%%%%%%%%%%%%%%%%%%%%%%%%%%%%%%%%%%%
\section{Discussion}
We have investigated the possible link between black-hole activity and star formation by probing star formation properties of large matched samples as a function of AGN power. Results of this work indicate that AGN and star formation are coupled to some degree for radio-quiet and radio-loud AGN, and also for different emission-line types of radio-loud AGN. A contrast between the SFRs and SSFRs of different types of active galaxies stands out in our results. Host galaxies of radio-quiet AGN have more stars forming than their radio-loud counterparts. In the case of the different emission-line classes, HERGs have higher SFR/SSFR than LERGs.

The galaxy masses are different for these objects and the influence of this should be taken into account. Therefore, a stacking analysis has also been carried out for radio-loud and radio-quiet AGN samples that are matched in galaxy mass. Mass-matched samples still indicate similar results; there is a slight increase in SFR of radio-quiet AGN as the AGN power increases. This rate is higher in radio-quiet AGN than HERGs and LERGs, but an increase in SFR and SSFR is also seen for HERGs and LERGs. We found a wide range of $\dot{M}_{\textrm{BH}}$ which are correlated with the SFRs of both samples. In the following section we interpret our results. 

\subsection{Interpretation of the correlation between SFR and AGN power}

In the simplest interpretation the correlation found between the rate of star formation and black-hole activity could be taken to be evidence for the synchronised growth of black holes and their host galaxies (Fig. \ref{l-q1}, \ref{l-q2}, \ref{mass-matched}, \ref{l-h}). This would then suggest that the existing cool gas supply on the host galaxy scale can feed the black hole in the centre of the galaxy at the same time as it allows for many more stars to form in the host galaxies of both radio-loud and radio-quiet active galaxies. In order to have this relationship, the available gas on host galaxy scales should be transported to the central regions by some physical mechanisms e.g. major mergers and secular processes. Major mergers are thought to be responsible for carrying a large amount of gas to the nuclear region of a galaxy \citep[e.g.][]{2005ref148,2010ref149}, thereby triggering powerful AGN and resulting in the co-evolution of black hole and its host galaxy in high accreting sources at high redshifts. Secular processes (large galaxy bars, disc instabilities, minor mergers etc.) are considered to be sufficient to move some of the available gas to the inner regions of galaxies to feed black holes \citep[e.g.][]{2011ref150}, especially in low accretion systems at low redshifts. Morphological studies of AGN host galaxies indicate that the correlations observed between the properties of black holes and that of their host galaxy are tight for ellipticals and bulge galaxies but not for disk galaxies with pseudo bulges \citep[e.g.][]{2008ref151}. The growth of ellipticals and bulge-dominated galaxies are commonly associated with mergers that can feed the most powerful quasars whereas low mass black holes have mostly been found in pseudobulges. The sources in our sample have low and moderate AGN powers at $z<0.6$. We do not have complete and detailed morphological studies of all sources but our rough analysis (Section 4.2) shows that most of the radio-loud AGN have elliptical hosts whereas radio-quiet AGN hosts tend to have more spiral characteristics. As mentioned above secular processes are often suggested to be responsible for fuelling disky (disk galaxies and pseudobulges), moderate luminosity galaxies in the low-redshift universe \citep[e.g.][]{2004ref159,2011ref160}. Considering all these implications for the galaxies in our sample, it is likely that internal processes are in many cases responsible for the formation of stars and driving the gas into the nuclear regions to feed black holes in their centre.

There is no clear answer to the question of when SFR correlates with AGN luminosity. However, the results of earlier studies indicate that a relation between AGN luminosity and SFR has been mainly observed for high-power sources (L$_{\textrm{AGN}}>$10$^{43}$ erg s$^{-1}$). Low-power (or quiescent galaxies) objects do not show such a relationship between these quantities in earlier works. For instance, \cite{2012ref104} found no significant correlation between galaxy-wide SFR and $\dot{M}_{\textrm{BH}}$ in local Seyferts (the mean distance is around 22 Mpc). \cite{2010ref128}, \cite{2010ref133} and \cite{2013ref71} used hard X-ray selected local Swift-BAT galaxies ($z<0.3$) and did not observe any correlation between L$_{\textrm{AGN}}$ and L$_{\textrm{SF}}$. Here in this work we also analyse a sample of low redshift AGN ($z<0.6$). However, our work is different from that of \cite{2013ref71} in that we take all components of different AGN types into account and calculate the total AGN power (by considering both mechanical and radiative output of AGN when it is present). We then derive average SFRs for our objects that are matched in total AGN power. Our analysis clearly shows that there is a relation between AGN power and SFR for mass matched samples of high power and low power AGN (Fig. \ref{l-q1}, \ref{mass-matched} and \ref{accretion}). This is a manifestation of a connection between AGN power and SFR that holds for low- and moderate-power sources at low redshifts as well. As mentioned above, \cite{2010ref128}, \cite{2010ref133} and \cite{2013ref71} have come to different conclusions: What leads to this difference? First of all, \cite{2010ref128} and \cite{2013ref71} used 60-$\mu$m far-IR luminosity, and \cite{2010ref133} used 870-$\mu$m submilimeter luminosity as a SFR indicator, while we derive the SFRs using the 250-$\mu$m luminosity, which should be minimally contaminated by AGN, old-stellar populations and torus. Another important difference is that the host galaxies of X-ray selected AGN may be different from the host galaxies of mid-IR or optical selected AGN \citep[e.g.][]{2009ref134,2013ref135}. Furthermore, the hard X-ray luminosity comes directly from the accretion disk whereas the [OIII] luminosity is derived from the extended narrow-line region of AGN. For this reason, the hard X-ray is expected to show more variation on short time scales (years) compared to the [OIII] luminosity, introducing scatter into any relationship based on the power derived from X-rays.

Another point to consider is the duty cycle of AGN. Some authors \citep[e.g.][]{2005ref43} have attempted to explain the discrepancy between the observed $\dot{M}_{\textrm{BH}}$-SFR relation and the expected one in luminous objects using a simple duty cycle model in which it is assumed that the growth of black holes occurs at high accretion rates at a fixed Eddington ratio and time, and they are off (accreting at unobservably low rates) otherwise. However, Fig. \ref{accretion} shows clearly that black holes can grow slowly at lower accretion rates. Low-power AGN have black holes growing slower than expected from the M$_{\textrm{BH}}$$-$M$_{\textrm{Bulge}}$ relationship and black holes in high-power AGN are growing faster. There is actually a continuous smooth growth of black holes instead of an on and off phases and, therefore, our results cannot be explained with a simple duty cycle scenario. It is worth noting that with our available data we cannot examine whether the available gas reservoir for star formation is in the galaxy bulge or arms. Therefore, it would be interesting to implement the same analysis using SFRs of de-composed bulge and galaxy disk.

Galaxy merger simulations have also predictions for the relation between SFR and AGN activity. In these models \citep[e.g.][]{2005ref119} it has been suggested that the SFR/$\dot{M}_{\textrm{BH}}$ ratio ranges between 200$-$600 at the peak of star formation and this value is expected to be around 1000 for low accretion rate objects ($\dot{M}_{\textrm{BH}}$=10$^{-2}$ M$_{\odot}$ yr$^{-1}$). Our results agree to some level with the predictions of this model. However, we have low and moderate luminosity galaxies with spiral (mostly in the radio-quiet AGN sample) and elliptical morphologies at low redshifts so many of the ellipticals in our sample may have ongoing mergers. \cite{2008ref120} used starburst disk models proposed by \cite{2005ref121} to study local AGN with hard X-ray luminosities of 10$^{43}$$-$10$^{44}$ erg s$^{-1}$ and they found that SFR/$\dot{M}_{\textrm{BH}}$ should be around $\sim$250 (for $\dot{M}_{\textrm{BH}}$ $\sim$ 0.3 M$_{\odot}$ yr$^{-1}$) for a disk with R$_{out}$=100 pc. This can only be observed for the whole galaxy of our radio-quiet AGN sample for black hole growth rates of $\dot{M}_{\textrm{BH}} = $ 2$\times$10$^{-3}$ $-$ 2$\times$10$^{-1}$ M$_{\odot}$ yr$^{-1}$. Recently, \cite{2014ref130} examined the correlation between $\dot{M}_{\textrm{BH}}$ and SFR of all galaxies for various AGN feedback models using the time evolution of a merger simulation and they found that although most feedback models produce the observed M-$\sigma$ relation, there are distinct differences between the results of various models. Some of these models result in predictions that are in qualitative agreement with our results reported here (see \cite{2013ref132} for details of these models.)

\subsection{Probable reasons for the difference between SFR/SSFR of radio-loud and radio-quiet AGN}
As mentioned in the previous section, radio-quiet and radio-loud AGN present a moderately strong relationship between their SFRs and AGN powers. If the observed trends are examined in detail it can be seen that for radio-quiet AGN at the high power end there is a tendency of constant SFR with increasing AGN power. This may be due to AGN feedback that starts having a significant effect on star formation when the gas supply reaches a certain level. In radio-loud AGN, black holes grow continuously. \cite{2009ref136} also reached similar conclusions. They analysed nearby galaxies selected from SDSS and found that the growth of black holes in local galaxies is present in two ways: if there is enough gas in the bulge the black hole can regulate itself; when the galaxy runs out of gas, the growth of black hole is regulated by the rate of mass loss due to evolved stars. 

Another question that arises from this work is: What could be the possible reason for the difference between the SFRs and SSFRs of radio-loud AGN and their radio-quiet cousins? The obvious difference between these AGN is the radio-emitting strong jets that we see in radio-loud AGN but not in their radio-quiet counterparts. If we especially focus on the results we have from the mass-matched samples where we eliminate the effect of mass, the striking difference in their SFRs may be suggesting that these strong jets are responsible for this difference, in the sense that there is the difference between the SFRs of radio-loud and radio-quiet AGN (lower/suppressed star formation in radio-loud AGN \footnote{The relation between radiative power and SFR could also be examined as a further test. This can only be done using HERGs and radio-quiet AGN since only these galaxies have radiative outputs. However, the number of HERGs is not large enough in our sample to carry out a statistically significant analysis. It can be seen in Figure \ref{mass-matched} that we have only two AGN power bins with both radio-quiet AGN and HERGs.}).

Another reason for this difference may be due to the difference between the morphology of galaxies in our mass sample. To check if we are comparing galaxies with similar morphologies, we searched the database of the Galaxy Zoo Project 2 \citep[GZ2;][]{2013ref106}. This was done by scanning the GZ2 catalogue sources with redshift\footnote{This redshift cut provides reliable classifications of galaxies, considering the SDSS image resolution.} $z<0.25$. Although we did not have classifications for all of our sources, this process gave an idea of their morphologies. We found that most radio-loud AGN are ellipticals but radio-quiet AGN have mixed morphologies (some of them are spirals and some are ellipticals). However, it should be pointed out that we cannot consider bulge and disk separately for radio-quiet AGN with our available data.

\subsection{The differences between SFR/SSFR in HERGs and LERGs}
We also had the advantage of exploring the star formation properties of large samples of emission-line classified radio galaxies using SFRs estimated by 250-$\mu$m $\textit{Herschel}$ luminosity matched in their AGN powers (Fig. \ref{l-h}). It is apparent that there is a relationship between the star formation and black hole accretion rates.

With regard to the comparison of SFRs and SSFRs for these galaxies, there is only very tentative evidence for a difference, in the contrast to the results of other studies \citep[e.g.][]{2013ref9,2010ref85,2008ref84}. Previously \cite{2013ref9} examined star formation properties of HERGs and LERGs by binning in radio luminosity and found a clear difference between these galaxies. It is important to note that in the present work we have matched our samples in total AGN power rather than luminosity, to compare SFRs of emission-line radio galaxies. Separating by radio luminosity (a proxy of the kinetic output) gives rise to a clear difference in the SFRs of HERGs and LERGs whereas this is less apparent when we take into account both kinetic and radiative outputs of HERGs. We cannot rule out a model in which the AGN power$-$SFR relation is the same for both radio-loud galaxy types, with the only difference being the way in which the accretion power manifests itself, though the HERG sample size in the current work is small.

 We also see a rise in SSFRs of both samples with AGN power. When stellar mass is taken into account the rising trend of SSFRs of LERGs is not as sharp as we see for their SFRs. This may be expected, as we know that LERGs are hosted by massive red ellipticals where most of the galaxy is dominated by an old stellar population, whereas HERGs tend to have lower mass and to be less evolved than LERGs. Other differences between these emission-line galaxies are the environment in which they are found \citep[e.g.][]{2004ref80,2008ref81} and the galaxy colours \citep[e.g.][]{2009ref82,2012ref83}. If the HERG/LERG difference is understood as an accretion rate switch \citep[][]{2012ref3, 2014ref50} then we would expect LERGs to have lower accretion rates and/or more massive black holes, implying smaller amounts of cold gas and/or more massive (hence more evolved) host galaxies.

\begin{table*}
\tabcolsep=0.07cm\begin{tabular}{cccccccccccccccc}
\hline
SDSS ID  & $z$ &      RA    &      DEC  &         Flux$_{250}$ &   Err$_{250}$  &  S$_{\mathrm{First}}$ &  Flux$_{\mathrm{[OIII]}}$ & Err$_{\mathrm{[OIII]}}$& SFR$_{\mathrm{H_{\alpha}}}$&    P16$_{\mathrm{SFR H_{\alpha}}}$& P84$_{\mathrm{SFR H_{\alpha}}}$& Mass & P16$_{\mathrm{Mass}}$  &   P84$_{\mathrm{Mass}}$ &    Type       \\
\hline
  588848899372744835 &   0.07 &  172.999  & -0.598 &   0.148 &  0.006  &  -  &      15.481  &   3.027  &    -0.310 & -1.050&         0.355 &   11.177 &   11.065  &  11.264   & Comp \\ 
  588848899372744859  &  0.11  &   173.011  &  -0.534  &   0.134  &   0.006   & -     &    12.503    &  2.158   &   0.488   &   0.119  &    0.796   &   11.009  &   10.886  &  11.114 &    Comp  \\
  588848899909746811 &   0.12  &  173.264   &   -0.154  &    0.020  &   0.006  &  -    &     14.536 &     2.664 &     0.183 &   -0.270   &  0.626  &   10.882 &   10.773   &    10.966   &  Comp   \\                 
.&.&.&.&.&.&.&.&.&.&.&.&.&.&.&\\
.&.&.&.&.&.&.&.&.&.&.&.&.&.&.&\\
.&.&.&.&.&.&.&.&.&.&.&.&.&.&.&\\
\hline
\end{tabular}
\caption[]{Full tables of measurements of the sample are available online from the journal website.} {}
\end{table*}

%%%%%%%%%%%%%%%%%%%%%%%%%%%%%%%%%%%%%%%%%%%%%%%%%%%%%%%%%%%%%%%%%%%%%%%%%%%%%
\section{Summary and Conclusions}
In this work, we have explored star formation properties, derived using 250-$\mu$m $\textit{Herschel}$ luminosity, of radio-quiet and radio-loud AGN samples and investigated whether they depend on black hole activity. We have also compared these features of radio-quiet and radio-loud AGN samples matched in their stellar mass and redshifts. The main results we have obtained are as follows.

\begin{itemize} 
\item{Examination of radio-loud and radio-quiet AGN samples matched only in their AGN powers shows that for both AGN samples the rate of star formation is increasing with increasing AGN activity. The same conclusion can be drawn for their relative galaxy growth rates (Fig. \ref{l-q1}).}

\item{A comparison of the star formation properties regarding these samples indicates that the host galaxies of radio-quiet AGN are forming more stars for a given black hole activity than their radio-loud cousins. This difference in the level of star formation per unit stellar mass goes up to an order of magnitude (Fig. \ref{l-q1}).}

\item{We also classified our radio-loud AGN sample in terms of emission-line type in order to compare their star formation properties. Both LERGs and HERGs present the same trend of increasing SFR with rising black-hole growth. In terms of their SSFRs both types of galaxies present a similar trend; their SSFRs are also increasing towards to higher AGN powers. They present almost a constant relative galaxy growth rate. There is only a marginal tendency for HERGs to have higher SFRs and SSFRs than LERGs when matched by AGN power (Fig. \ref{l-h}).}

\item{To account for the likely influence of stellar mass we composed radio-loud and radio-quiet AGN samples matched in their galaxy masses. The same stacking analysis has been implemented for these samples. Our findings suggest that the black-hole growth and the star formation rate are coupled. The amount of stars forming for a given time increases with increasing AGN power for both AGN samples (Fig. \ref{mass-matched}).}

\item{When we take into account the effect of intrinsic correlation between redshift and SFR in our stacking analysis we have shown that the strong correlation between SFR and AGN power for radio-loud and radio-quiet AGN sample does not vanish. Furthermore, we have evaluated the effect of SFR evolution with redshift. This analysis showed that the difference between SFRs of radio-loud and radio-quiet AGN is not due to this effect.}

\item{We have also assessed the relation between SFR and $\dot{M}_{\textrm{BH}}$ and found that both radio-quiet and radio-loud AGN have a range of ratios of SFR/$\dot{M}_{\textrm{BH}}$. A comparison of our results with the observed correlation suggests that low-power AGN have black holes growing slower than expected from the M$_{\textrm{BH}}$$-$M$_{\textrm{Bulge}}$ relationship and black holes in high-power AGN are growing faster. Our results cannot be explained with a simple model of duty cycle of AGN (Fig. \ref{accretion}).}

\item{Reasons behind the apparent difference in SFR of radio-quiet AGN and HERGs and LERGs have been explored. This difference is still seen when comparing the samples of radio-loud and radio-quiet AGN matched in their stellar mass (Fig. \ref{mass-matched}). There may be two possible reasons for this. Either the strong jets we observe in radio-loud AGN suppress star formation in these galaxies or the difference between the galaxy morphology of the AGN samples leads to this observed disparity, or some combination of the two. Nevertheless, a direct role for feedback from the radio jets is not ruled out.}

\end{itemize}

\section*{Acknowledgements}
GG thanks the University of Hertfordshire for a PhD studentship. We would like to thank Gianfranco De Zotti and Michal Michalowski for their useful comments. LD, RJI and SJM acknowledge support from the European Research Council Advanced grant COSMICISM. NB acknowledges support from the EC FP7 SPACE project ASTRODEEP (Ref.No: 312725). This work has made use of the University of Hertfordshire Science and Technology Research Institute high-performance computing facility. This publication makes use of data products from the Wide-field Infrared Survey Explorer, which is a joint project of the University of California, Los Angeles, and the Jet Propulsion Laboratory/California Institute of Technology, funded by the National Aeronautics and Space Administration. \textit{Herschel}-ATLAS is a project with \textit{Herschel}, which is an ESA space observatory with science instruments provided by European-led Principal Investigator consortia and with important participation from NASA. The H-ATLAS website is http://www.h-atlas.org/. The GMRT input catalogue is based on data takes from the SDSS and the UKIRT Infrared Deep Sky Survey.

\bibliographystyle{mn2e}
\bibliography{ref-sfr-agn}

\label{lastpage}

\end{document}